\documentclass{jfm}
\usepackage{graphicx}
\usepackage{epstopdf, epsfig}
\usepackage{tikz}
\usepackage{amsmath}
\usepackage{physics}
\usepackage{mathrsfs}
\usepackage{amsmath}
\usepackage{tikz}
\usepackage{mathdots}
\usepackage{yhmath}
\usepackage{cancel}
\usepackage{color}
\usepackage{siunitx}
\usepackage{array}
\usepackage{bm}
\usepackage[normalem]{ulem}
\usepackage{multirow}
\usepackage{amssymb}
\usepackage{gensymb}
\usepackage{tabularx}
\usepackage{extarrows}
\usepackage{booktabs}
\usetikzlibrary{fadings}
\usetikzlibrary{patterns}
\usetikzlibrary{shadows.blur}
\usetikzlibrary{shapes}

\usepackage[normalem]{ulem}
\usepackage[colorlinks=true,linkcolor=blue,urlcolor=blue,citecolor=blue]{hyperref}

\shorttitle{Forced and natural thermal convection in a cylindrical layer at low $Pr$ number}
\shortauthor{F. Rein, L. Car\'enini, F. Fichot, B. Favier, M. Le Bars}

\title{Interaction between forced and natural convection in a thin cylindrical fluid layer at low Prandtl number}

\author{F. Rein \aff{1,2}
  \corresp{\email{florian.rein@protonmail.com}},
   L. Car\'enini\aff{2}, F. Fichot\aff{2}, B. Favier\aff{1}, M. Le Bars\aff{1}}

\affiliation{\aff{1}Aix Marseille Univ, CNRS, Centrale Marseille, IRPHE, Marseille, France \aff{2} IRSN, St Paul lez Durance, France}

\begin{document}

\maketitle

\begin{abstract}
Motivated by nuclear safety issues, we study the heat transfers in a thin cylindrical fluid layer with imposed fluxes at the bottom and top surfaces (not necessarily equal) and a fixed temperature on the sides.
We combine Direct Numerical Simulations and a theoretical approach to derive scaling laws for the mean temperature and for the temperature difference between the top and bottom of the system.
We find two asymptotic scaling laws depending on the flux ratio between the upper and lower boundaries.
The first one is controlled by heat transfer to the side, for which we recover scaling laws characteristic of natural convection \citep{Batchelor}.
The second one is driven by vertical heat transfers analogous to Rayleigh-Bénard convection \citep{GL}.
We show that the system is inherently inhomogeneous, and that the heat transfer results from a superposition of both asymptotic regimes.
Keeping in mind nuclear safety models, we also derive a one-dimensional model of the radial temperature profile based on a detailed analysis of the flow structure, hence providing a way to relate this profile to the imposed boundary conditions.
\end{abstract}


\section{Introduction}
\input{Fig_tex/fig_intro}
When a severe accident (SA) occurs in a nuclear power plant, the radioactive fuel and reactor metallic components melt and form a fluid called corium.
The corium relocates from the core to the lower plenum of the reactor vessel, where non-miscible oxidic and metallic phases separate: the oxide phase contains the majority of the decay heat from radioactive elements and  heats from below the less dense liquid metal phase floating at the surface (Figure \ref{intro}\textcolor{blue}{$a$}).
This top metal layer is usually thinner compared to the oxide layer and thus concentrates the power from the oxide to the vessel wall. This phenomenon is often referred to as the ``focusing effect" in the nuclear safety litterature.
When the external cooling of the reactor vessel is implemented as a SA management strategy \citep{THEOFANOUS,IVR}, addressing the heat transfer through the top metal layer is fundamental to predict the failure of the vessel or to justify its integrity. 
This work focuses on this issue: a liquid metal layer considered as a cylindrical layer heated from below, cooled at the side with a constant temperature (assuming that the wall is being ablated and thus maintained at its melting temperature) and cooled at the top by radiative heat transfer.
One difficulty of the problem lies in the top boundary condition, due to the interdependence of variables: the radiative heat flux from the upper surface depends on the surface temperature, which is affected by heat transfer within the metal layer, which in turn depends on the efficiency of the radiative heat flux.
The approach proposed is to prescribe a uniform heat flux leaving from the top of the layer and focus on analysing the temperature profiles, fluid behavior and heat transfers. This will allow  correlating these outputs with the input control parameters and encompass all possible configurations within the reactor.
Indeed, depending on the height of the metal layer and on the state of the reactor vessel structures above the pool, the radiative heat transfer can either play a major role for the power dissipation from the metal or be negligible. This approach has also the advantage of decoupling the study of the metal layer from the modelling of the radiative heat transfer. As illustrated in \cite{LeGuennic}, the consideration of the top radiative heat transfer introduces assumptions on its modelling directly in correlations esthablished for the behavior of the metal layer. With the present approach, coupling with more detailed radiative heat transfer models will be possible \citep{rein}. Impact of considering a uniform heat flux compared to a more realistic radiative exchange will be evaluated in future studies.

Given the specific boundary conditions, a mixture of different types of convection can be expected.
Bottom heating and top cooling is reminiscent of Rayleigh-B\'enard configurations with an imposed flux often investigated in the literature \cite[e.g.][]{,hurlee,cp,doering_2002,verzicco_2008,johnston,fantuzzi_2018}.
The lateral cooling is additionally expected to sustain natural convection, which has also been the subject of numerous studies \cite[e.g.][]{Batchelor,churchill,bejan,GEORGE,wells_worster_2008,Ng_2015,shishkina}.
In integral SA codes, like the ASTEC code developed by IRSN \citep{CHATELARD2014}, the entire process of a reactor core meltdown accident is simulated, from initiating events to the release of radioactive materials. Different modules address different aspects of the accident, the corium behaviour in the lower plenum of the vessel being one of them \citep{Fdc-ASTEC}. In such code, the focusing effect evaluation is based on a simplified approach proposed by \citep{THEOFANOUS}, which combines correlations from  both Rayleigh-B\'enard and natural convection .
It is assumed that the fluid in the bulk is thoroughly mixed and that the vertical heat transfer is symmetrical, meaning that the temperature difference between the bottom and the bulk is equal to the temperature difference between the bulk and the top.
The validity of this approach was checked, in particular with the BALI-Metal facility \citep{bali}. Water was used to simulate the corium, and the top boundary condition was controlled by conduction through an epoxy plate and a temperature-regulated heat exchanger. This setup was designed to closely resemble the conditions in a reactor with radiative heat transfer at the top. BALI-Metal tests have shown that, for a shallow layer thickness (aspect ratio above 10), the \textit{0D} model overestimates the side heat flux. However, CFD simulations of the metal layer \citep{IVMR} showed that fluid properties (water vs. steel) have a significant effect on the global behavior, especially the Prandtl number. With steel, the lateral heat flux is up to $50\%$ higher  compared to water under similar conditions. This questions the validity of using water as a simulant for molten steel, and consequently the previously derived model for integral SA codes.

More generally, the competition between forced convection involving dominant vertical heat fluxes and horizontal or natural convection involving dominant horizontal heat fluxes is at the core of many geophysical situations.
The competition between Rayleigh-B\'enard and horizontal modes of convection is important for the dynamics within subglacial lakes in Greenland and Antarctica \citep{couston2022competition,Livingstone2022}.
Planetary oceans are another example, since they receive latitudinally-dependent solar radiations while being heated from the bottom by the geothermal flux \citep{Wang2016}.
Finally, heterogeneous heat fluxes along the core-mantle boundary in the Earth's core, which are due to large-scale convective patterns within the solid mantle, can sustain large-scale azimuthal flows \citep{Sumita1999,Mound2017}.

This paper presents 3D Direct Numerical Simulations (DNS) of a liquid layer with a fixed Prandtl number of $0.1$ motivated by nuclear safety issues involving liquid metals \citep{IVR}, with the aim of: (i) getting a better understanding of the fluid behavior and heat transfers for different characteristics found in reactors, (ii) establishing scaling laws, and (iii) deriving a one-dimensional model suitable for use in integral SA codes.

This article is divided into four sections. The first section introduces the governing equations and the numerical simulation tool used. The second section focuses on identifying the heat transfer mechanisms for the asymptotic regimes (dominant side or top heat flux) by analysing scaling laws for mean temperature variables. We conclude that \textit{a minima}, a 1D radial temperature description of the turbulent regime is necessary for nuclear safety evaluations. The third section hence delves into the fluid flow structure to determine this temperature profile. Finally, the last section discusses the implications of our results for nuclear safety evaluations and outlines future developments of our work.

\section{Mathematical and numerical formulation}\label{sec:rules_submission}
\subsection{Governing equations}

We consider the flow of an incompressible fluid with buoyancy effects being included using the Boussinesq approximation.
The fluid is confined within a cylinder of thickness $H$ and radius $R$ (see Figure \ref{intro}\textcolor{blue}{b}) and gravity is pointing downward $\bm{g}=-g\bm{e}_z$.
It is heated from below with a uniform heat flux per unit area $\phi_{\textrm{in}}$ and cooled from above by a uniform outgoing flux $\phi_{\textrm{out}}$.
Note that we are interested in the cases where $\phi_{\textrm{in}}\neq\phi_{\textrm{out}}$ so that the residual heat flux is necessarily escaping the domain through the side boundary.
The dimensional temperature on the side boundary is fixed at $\theta_0$. We model the bottom interface between the oxide layer and the liquid metal layer by a no-slip rigid boundary \cite[a rigid crust forms at the oxide surface due to cooling,][]{IVR}; we model the upper free surface of the liquid layer by a rigid stress-free boundary, neglecting free surface deformations. The side boundary is a rigid no slip boundary. Lengths are rescaled using the height of the cylinder $H$ while time is rescaled using the vertical diffusive timescale $H^2/\kappa$, with $\kappa$ the constant thermal diffusivity.
The dimensionless temperature $T$ is defined relatively to the imposed side temperature and rescaled using the imposed bottom flux $\phi_{\textrm{in}}$
\begin{equation}
\label{ndt}
    T=\frac{k}{\phi_{\textrm{in}}H}\left(\theta-\theta_0\right) \ ,
\end{equation}
where $k$ is the thermal conductivity.
The dimensionless conservation equations of momentum, mass and energy are then
\begin{equation}
\begin{aligned}
\frac{1}{Pr}\left(\frac{\partial\bm{u}}{\partial t}+ \bm{u\cdot\nabla}\bm{u}\right) =-\bm{\nabla}P +Ra_{\phi} T\bm{e}_z+\bm{\nabla}^2\bm{u} \ ,
\end{aligned}
\label{qdm}
\end{equation}
\begin{equation}
\bm{\nabla\cdot u}=0 \ ,
\label{m}
\end{equation}
\begin{equation}
\frac{\partial T}{\partial t}+ \bm{u}\cdot\bm{\nabla} T=\nabla^2 T \ .
\label{nrj}
\end{equation}
$\bm{u}$, $P$ and $T$ are the dimensionless velocity, pressure and temperature of the fluid respectively.
The problem is characterised by four dimensionless parameters:
the aspect ratio $\Gamma$, the flux ratio $R_F$, the Rayleigh-Roberts number $Ra_{\phi}$ which is based on the heat flux imposed at the bottom $\phi_{\textrm{in}}$, and the Prandtl number $Pr$ fixed to $0.1$ throughout the paper. They are defined by 
\begin{equation}
\Gamma = \frac{R}{H},~~~R_F=\frac{\phi_{\textrm{out}}}{\phi_{\textrm{in}}},~~~Ra_{\phi}=\frac{\beta g \phi_{\textrm{in}} H^4}{k \nu \kappa},~~~Pr=\frac{\nu}{\kappa}=0.1 \ ,
\end{equation}
where $\beta$ is the thermal expansion coefficient and $\nu$ is the kinematic viscosity, both assumed to be constant.
The dimensionless boundary conditions can be written as

\begin{equation}
\begin{aligned}
\bm{u}(z=0)=\bm{0}~~~\mbox{and}~~~\frac{\partial T}{\partial z}(z=0)=-1 \ ,\\
\frac{\partial u}{\partial z}(z=1)=\frac{\partial v}{\partial z}(z=1)=w(z=1)=0~~~\mbox{and}~~~\frac{\partial T}{\partial z}(z=1)=-R_F \ ,\\
\bm{u}(r=\Gamma)=\bm{0}~~~\mbox{and}~~~ T(r=\Gamma)=0 \ ,
\end{aligned}
\label{bcn}
\end{equation}
where $\bm{u}=\left(u,v,w\right)$ are the velocity components in cylindrical coordinates $\bm{e}_r,\bm{e}_{\varphi},\bm{e}_z$ respectively.

\subsection{Numerical approach}\label{sec:types_paper}

\subsubsection{Nek5000}

The governing equations \eqref{qdm}-\eqref{nrj} with boundary conditions \eqref{bcn} are solved numerically using \href{https://nek5000.mcs.anl.gov/}{Nek5000} \citep{Fischer1997,Deville2002}, which has been used extensively in thermal convection studies \cite[e.g.][]{Scheel2013,Leard2020,terrien2023suppression}.
The entire cylindrical geometry is discretised using up to $\mathcal{E}=36608$ hexahedral elements which have been refined close to all boundaries to properly resolve viscous and thermal boundary layers.
The velocity is discretised within each element using Lagrange polynomial interpolants based on tensor-product arrays of Gauss–Lobatto–Legendre quadrature points.
The polynomial order $N$ on each element varies between $6$ and $10$ in this study.
We use the $3/2$ rule for dealiasing with extended dealiased polynomial order $3N/2$ to compute nonlinear products.
A third-order time stepping using a mixed explicit-implicit backward difference approach is used.
A summary of the simulations physical and numerical parameters is provided in table \ref{table:1} in Appendix \ref{appA}. 

\subsubsection{Numerical protocol and statistics}  

We initialise all simulations with a fluid at rest and a uniform temperature field $T=0$ everywhere.
Infinitesimal temperature perturbations of amplitude $10^{-3}$ are introduced.
Thermal convection grows during a transient which typically lasts for approximately $5$ vertical diffusive times, and which is longer as the aspect ratio $\Gamma$ increases.
Once the system has reached a statistically-stationnary state, various spatio-temporal averages are computed.
Note that we have tested that different initial conditions (for example starting with the equilibrium diffusive temperature distribution) eventually lead to the same statistically-stationnary state. 

We first define the temporal and volume average operator $\left<.\right>$ over the whole fluid domain volume $V$ and over time $\tau$ as
\begin{equation}
\left<T\right>=\frac{1}{\tau V}\int_{t_0}^{t_0+\tau}\int_{V}T\mathrm{d}V\mathrm{d}t \ .
\label{avg}
\end{equation}
The typical $\tau$ value ranges between $2$ and $0.2$ diffusive times for the lowest and the largest Rayleigh-Roberts numbers respectively. Additionally, adding specific variables as a subscript means that an average along those specific directions is made.
We always consider temporal averages during the statistically steady state so that the time variable is never explicitely written.
For instance, $\left<.\right>_{\varphi}$ indicates an average in time and along the azimuthal direction only.

\subsubsection{Filtered simulations}

While most of the results discussed below are obtained using Direct Numerical Simulations (DNS), some extreme cases were only accessible via filtered simulations following the approach described in \cite{fischer}.
To distinguish between DNS and filtered simulations, a viscous dissipation criterion has been used. 
The mean viscous dissipation rate $\epsilon$, is defined by  
\begin{equation}
\epsilon=2Pr\left< \bm{S} :\bm{S}\right>~\mbox{with}~\bm{S}=\frac{1}{2}\left(\bm{\nabla u} + \bm{\nabla u}^T \right).
\end{equation}
A simulation with polynomial order $N$ is considered to be a DNS when the time and volume averaged viscous dissipation $\epsilon$ is varying by less than $5\%$ when compared with the same simulation but using $N+2$ polynomial order.
A simulation failing to satisfy this criteria is labelled as filtered and numerical stability is ensured by using a $1\%$ filter on the last 2 polynomials \citep{fischer}.
Notice that in the system studied, the Prandtl is $0.1$, meaning that the filter does not impact the more diffusive temperature field but only the viscous dissipation at small scales.

Alternatively, we also used the criterion discussed in \cite{Scheel2013} which compares the isotropic Kolmogorov dissipative scale with the numerical grid size.
For all the DNS simulations presented in this study, the numerical grid size ($L$) is below the Kolmogorov dissipative scale ($\eta_K$).

\section{Results}
\begin{figure}
   \centering
    \includegraphics[scale=0.20]{
    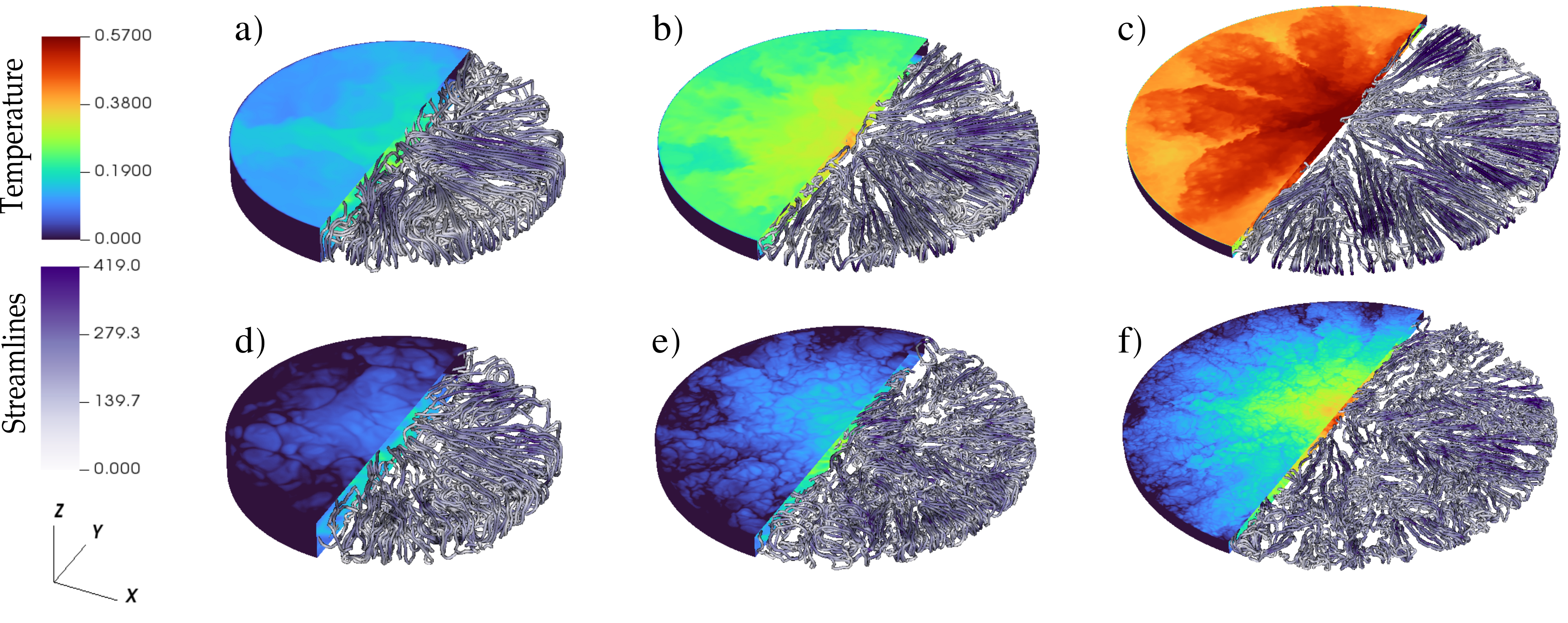}
   \caption{Each of the 6 subplots shows the 3D view of (i) on the left half, the temperature field at the top surface and (ii) on the right half, some velocity streamlines. The top row shows results for $R_F=0.1$ and  the bottom row for $R_F=0.9$, with $\Gamma=4$ for $a)$ and $d)$, $\Gamma=8$ for $b)$ and $e)$, and $\Gamma=16$ for $c)$ and $f)$. In all cases, $Pr=0.1$ and $Ra_{\phi}=10^7$. Note that the temperature color scale is the same for all plots.}
   \label{low_ra}
\end{figure}

\subsection{Qualitative overview}

Let us start with a qualitative description of the different flow regimes.
First, irrespective of the control parameters, no motion-less steady state exists in this system.
Maintaining a constant temperature at the side generates a radial temperature gradient, which cannot be balanced by the hydrostatic pressure gradient.
This leads to natural convection in the form of a downward recirculation along the side boundary and then towards the center along the bottom boundary.
At low $Ra_{\phi}$, this flow is axisymmetric, while at larger $Ra_{\phi}$, an instablity develops and breaks the symmetry, leading to a drifting thermal branches pattern associated to the most extreme heat fluxes and temperatures found in the system.
While these observations call for a linear stability analysis to identify the symmetry-breaking mechanism, we leave this aspect to future works.
The focus of the present work is to identify scaling laws in the turbulent regime at large Rayleigh-Roberts number $Ra_{\phi}$, irrespective of the underlying linear instability mechanism.

We first present three-dimensional visualisations showing both the temperature field at the surface and streamlines colored with the velocity amplitude in Figure \ref{low_ra}.
We focus on the representative case $Ra_{\phi}=10^7$ and compare two flux ratios, $R_F=0.1$ and $0.9$, and three aspect ratios, $\Gamma=4$, $8$ and $16$.
At low flux ratio (see Figures \ref{low_ra}(a), (b) and (c)), a large-scale temperature pattern is clearly visible, with a number of azimuthal branches increasing with $\Gamma$. This thermal pattern is associated with intense radially outward flows.
As we will see, these regimes are dominated by convective motions reminiscent of horizontal convection \citep{Hughes2008}.
At high flux ratio (see Figures \ref{low_ra}(d), (e) and (f)), the system is more azimutally symmetrical and more intense fluctuations cover most of the domain.
We also observe a clear temperature gradient between the core of the cylinder and the isothermal side boundary.
As we will see, these regimes are dominated by thermal structures reminiscent of Rayleigh-B\'enard convection \citep{bodenschatz2000recent}.

\subsection{Low flux ratio regime\label{lowrf}}

This section is devoted to the low flux ratio regime for which we fix $R_F=0.1$ as a representative value.
We first discuss the mean temperature scaling as a function of the two other input parameters $Ra_{\phi}$ and $\Gamma$ before providing a theoretical explanation based on simple dimensional arguments.

\subsubsection{Scaling for $R_F=0.1$}

In this section we fixed $R_F=0.1$, meaning that $90\%$ of the power transferred through the lower boundary goes out through the side, and we seek scaling laws for the mean temperature systematically varying $Ra_{\phi}$ and $\Gamma$.
For each simulation, the statistically stationary state is reached, which typically takes $5$ diffusive timescales, and we compute the mean temperature of the system noted $\left<T\right>$ using the space-time average operator defined in \eqref{avg}.

\begin{figure}
   \centering
    \includegraphics[scale=0.32]{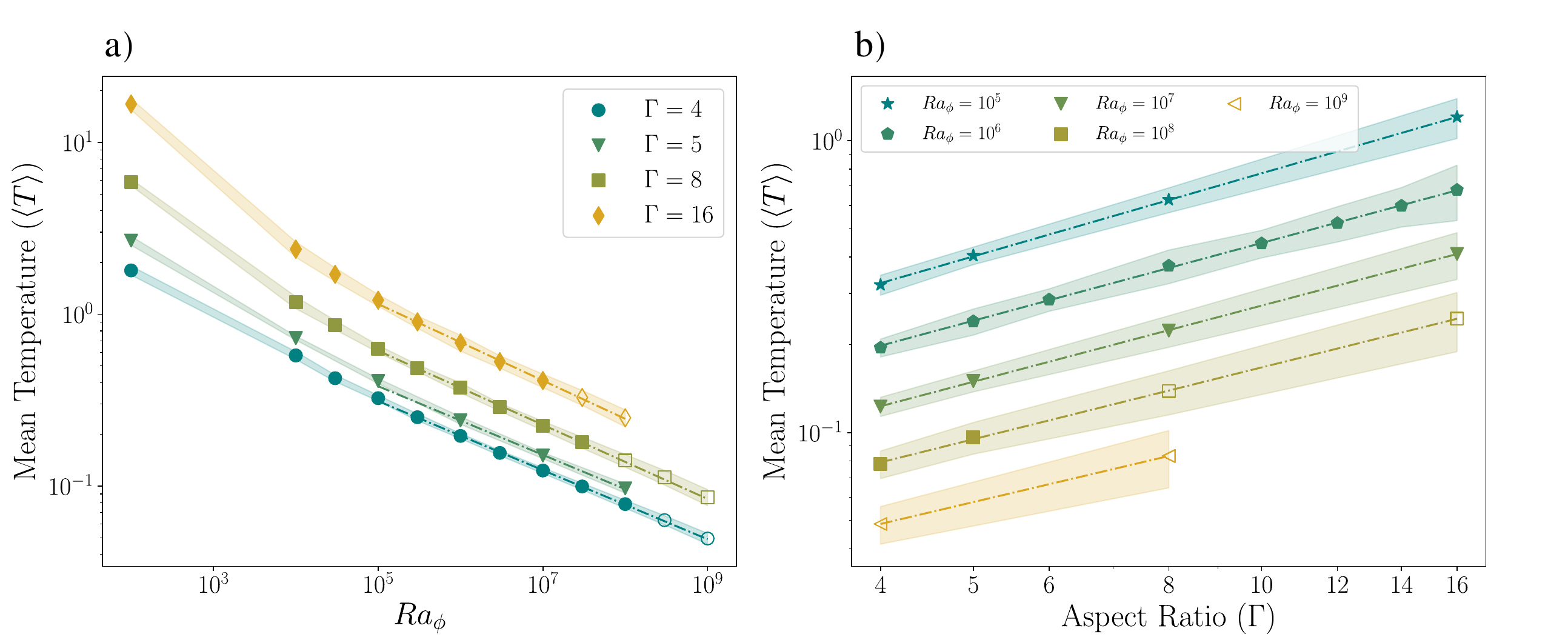}

   \caption{Log-log plot of the mean temperature evolution with Rayleigh number for different aspect ratios in a) and with the aspect ratio for different Rayleigh numbers in b). The width of the thick light lines is equal to $3$ times the standard deviation of the mean temperature time series at the statistically stationary state. The dash-dotted lines show the best power-law fit for each $\Gamma$ in a) and $Ra_{\phi}$ in b). Empty/full symbols indicate respectively filtered/DNS simulations. $Pr=0.1$ and $R_F=0.1$.}
   \label{tm_ra}
\end{figure}

The mean temperature evolution with $Ra_{\phi}$ and $\Gamma$ is reported in Figure \ref{tm_ra}.
When the value of $Ra_{\phi}$ increases, the mean non-dimensional temperature of the system decreases as expected (see Figure~\ref{tm_ra}(a)).
This is because a higher Rayleigh-Roberts number results in more efficient heat transfers within the system, leading to an average temperature getting closer to the side temperature which is zero in our dimensionless units \eqref{ndt}.
Note however that the dimensional temperature obviously increases when we increase the heat flux.
When $Ra_{\phi}$ exceeds $10^5$, a power law behavior emerges with an exponent that appears to be unaffected by $\Gamma$.
Upon estimating the power exponent (best fit using the least square method), it has been found that $\left<T\right>\sim Ra_{\phi}^{-0.20\pm 0.03}$. To determine the mean exponent, we computed the average of the exponents associated with $\Gamma=4,5,8,16$ considering $Ra_{\phi} \geq 10^5$ data only, while the variability is quantified by the largest difference between these exponents. This scaling law is derived by considering DNS and filtered simulation data. The exclusion of the filtered data from the analysis only leads to a slight alteration in the scaling law, resulting in $\left<T\right>\sim Ra_{\phi}^{-0.21\pm 0.03}$. 

Additionally, when $\Gamma$ increases at constant $Ra_{\phi}$, the mean temperature of the system increases.
Indeed, when $\Gamma$ increases for a fixed $Ra_{\phi}$,
the ratio between the heating bottom surface and the cooling lateral surface also increases, leading to a larger global energy input into the system. Figure \ref{tm_ra}(b) shows that when $Ra_{\phi}$ is greater than $10^5$, a power law behavior in $\Gamma$ also emerges. Considering both DNS and filtered simulation data, and employing a consistent methodology for exponent estimation as applied to the $Ra_{\phi}$ dependence, we obtain $\left<T\right>\sim \Gamma^{0.83 \pm 0.11}$ whereas we find $\left<T\right>\sim \Gamma^{0.89 \pm 0.08}$ without taking filtered data into account. Once again these results indicate a minor influence of the filtered simulations on the overall scaling behavior.

The two scalings can be combined leading to the following final power law
\begin{equation}
\left<T\right> \sim Ra_{\phi}^{-1/5}~\Gamma^{4/5} \ ,
\label{scal}
\end{equation}
where the particular exponent values will be theoretically justified below (see section~\ref{sec:theorf01}) and fall well into the fitted range of exponents found from our simulations. 
\begin{figure}
   \centering
    \includegraphics[scale=0.32]{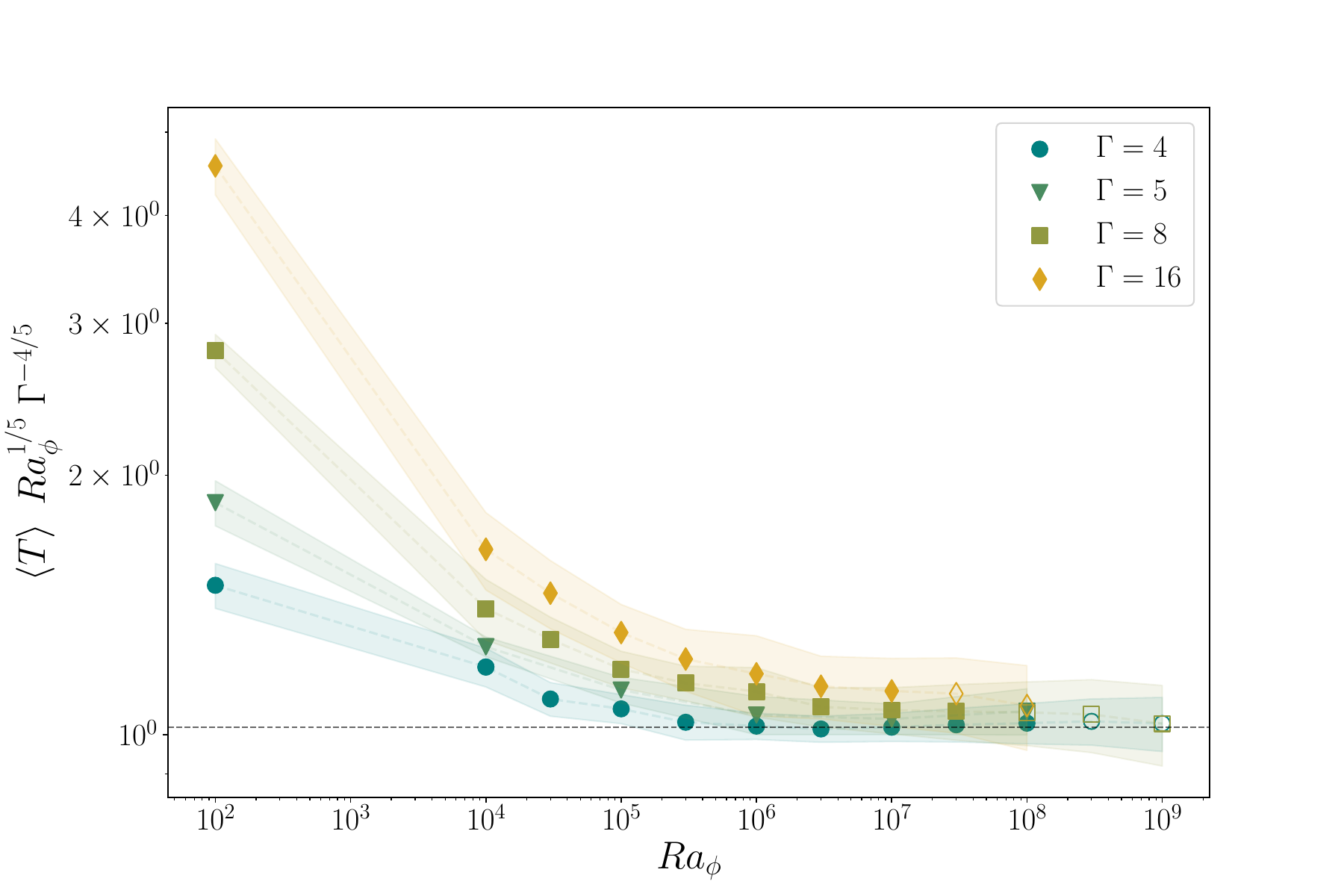}
   \caption{Compensated mean temperature following \eqref{scal} as a function of $Ra_{\phi}$. Different symbols/colors correspond to different aspect ratios. The grey areas indicate $3$ times standard deviations of the mean temperature time series at the statistically stationary state multiplied by $Ra_{\phi}^{1/5}\Gamma^{-4/5}$. Empty/full symbols indicate respectively filtered/DNS simulations. $Pr=0.1$ and $R_F=0.1$}
   \label{fignu}
\end{figure}
Figure \ref{fignu} shows the mean temperature, compensated by scaling \eqref{scal} as a function of $Ra_{\phi}$.
All rescaled data converge to the same constant close to unity.
For $\Gamma=4$, the scaling law seems to apply when $Ra_{\phi}>10^5$, whilst it is necessary to wait until $Ra_{\phi}=10^8$ when $\Gamma=16$.
In the following section, we use dimensional analysis to gain insight into the physics underlying this scaling.

\subsubsection{Theoretical analysis for $R_F=0.1$\label{sec:theorf01}}

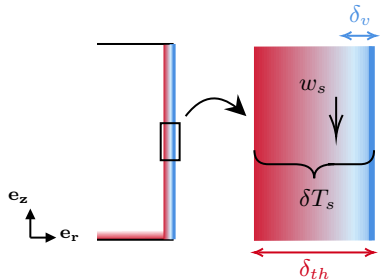
\begin{figure}
    \centering

  
\tikzset {_kd8la2hga/.code = {\pgfsetadditionalshadetransform{ \pgftransformshift{\pgfpoint{0 bp } { -3.5 bp }  }  \pgftransformrotate{-269 }  \pgftransformscale{2 }  }}}
\pgfdeclarehorizontalshading{_f80fwz9q8}{150bp}{rgb(0bp)=(0.82,0.01,0.11);
rgb(37.5bp)=(0.82,0.01,0.11);
rgb(57.23214285714286bp)=(1,1,1);
rgb(100bp)=(1,1,1)}

  
\tikzset {_cp8qz2mi4/.code = {\pgfsetadditionalshadetransform{ \pgftransformshift{\pgfpoint{0 bp } { 0 bp }  }  \pgftransformrotate{0 }  \pgftransformscale{2.6 }  }}}
\pgfdeclarehorizontalshading{_ihtuodsle}{150bp}{rgb(0bp)=(0.82,0.01,0.11);
rgb(37.5bp)=(0.82,0.01,0.11);
rgb(56.160714285714285bp)=(0.82,0.92,0.98);
rgb(62.5bp)=(0.29,0.56,0.89);
rgb(100bp)=(0.29,0.56,0.89)}

  
\tikzset {_40kxyhx9w/.code = {\pgfsetadditionalshadetransform{ \pgftransformshift{\pgfpoint{0 bp } { 0 bp }  }  \pgftransformrotate{0 }  \pgftransformscale{2.6 }  }}}
\pgfdeclarehorizontalshading{_b53qhprlw}{150bp}{rgb(0bp)=(0.82,0.01,0.11);
rgb(37.5bp)=(0.82,0.01,0.11);
rgb(53.39285714285714bp)=(0.82,0.92,0.98);
rgb(62.5bp)=(0.29,0.56,0.89);
rgb(100bp)=(0.29,0.56,0.89)}
\tikzset{every picture/.style={line width=0.75pt}} 

\begin{tikzpicture}[x=0.75pt,y=0.75pt,yscale=-1,xscale=1]

\draw  [draw opacity=0][shading=_f80fwz9q8,_kd8la2hga] (80.53,9704.73) -- (120.9,9704.73) -- (120.9,9711.39) -- (80.53,9711.39) -- cycle ;
\draw  [draw opacity=0][shading=_ihtuodsle,_cp8qz2mi4] (163.78,9711.95) -- (163.78,9614.91) -- (222.89,9614.71) -- (222.89,9711.75) -- cycle ;
\draw   (79.2,9613.61) -- (122.98,9613.61) -- (122.98,9711.39) -- (79.2,9711.39) -- cycle ;
\draw [color={rgb, 255:red, 74; green, 144; blue, 226 }  ,draw opacity=1 ]   (210.32,9609.51) -- (222.97,9609.51) ;
\draw [shift={(224.97,9609.51)}, rotate = 180] [color={rgb, 255:red, 74; green, 144; blue, 226 }  ,draw opacity=1 ][line width=0.75]    (4.37,-1.96) .. controls (2.78,-0.92) and (1.32,-0.27) .. (0,0) .. controls (1.32,0.27) and (2.78,0.92) .. (4.37,1.96)   ;
\draw [shift={(208.32,9609.51)}, rotate = 0] [color={rgb, 255:red, 74; green, 144; blue, 226 }  ,draw opacity=1 ][line width=0.75]    (4.37,-1.96) .. controls (2.78,-0.92) and (1.32,-0.27) .. (0,0) .. controls (1.32,0.27) and (2.78,0.92) .. (4.37,1.96)   ;
\draw [color={rgb, 255:red, 0; green, 0; blue, 0 }  ,draw opacity=1 ]   (205,9640) -- (205,9658) ;
\draw [shift={(205,9660)}, rotate = 270] [color={rgb, 255:red, 0; green, 0; blue, 0 }  ,draw opacity=1 ][line width=0.75]    (10.93,-3.29) .. controls (6.95,-1.4) and (3.31,-0.3) .. (0,0) .. controls (3.31,0.3) and (6.95,1.4) .. (10.93,3.29)   ;
\draw [color={rgb, 255:red, 74; green, 144; blue, 226 }  ,draw opacity=1 ][line width=2.25]    (122.98,9613.61) -- (122.98,9711.13) ;
\draw  [draw opacity=0][shading=_b53qhprlw,_40kxyhx9w] (118.4,9711.15) -- (118.4,9614.02) -- (122.98,9614) -- (122.98,9711.13) -- cycle ;
\draw [color={rgb, 255:red, 208; green, 2; blue, 27 }  ,draw opacity=1 ]   (165.94,9717.74) -- (221.89,9717.74) ;
\draw [shift={(223.89,9717.74)}, rotate = 180] [color={rgb, 255:red, 208; green, 2; blue, 27 }  ,draw opacity=1 ][line width=0.75]    (4.37,-1.96) .. controls (2.78,-0.92) and (1.32,-0.27) .. (0,0) .. controls (1.32,0.27) and (2.78,0.92) .. (4.37,1.96)   ;
\draw [shift={(163.94,9717.74)}, rotate = 0] [color={rgb, 255:red, 208; green, 2; blue, 27 }  ,draw opacity=1 ][line width=0.75]    (4.37,-1.96) .. controls (2.78,-0.92) and (1.32,-0.27) .. (0,0) .. controls (1.32,0.27) and (2.78,0.92) .. (4.37,1.96)   ;
\draw [color={rgb, 255:red, 74; green, 144; blue, 226 }  ,draw opacity=1 ][line width=2.25]    (222.89,9614.71) -- (222.89,9711.75) ;
\draw  [draw opacity=0][fill={rgb, 255:red, 255; green, 255; blue, 255 }  ,fill opacity=1 ] (77.09,9610) -- (85,9610) -- (85,9713.33) -- (77.09,9713.33) -- cycle ;
\draw    (51.61,9710.32) -- (61.63,9710.32) ;
\draw [shift={(64.63,9710.32)}, rotate = 180] [fill={rgb, 255:red, 0; green, 0; blue, 0 }  ][line width=0.08]  [draw opacity=0] (5.36,-2.57) -- (0,0) -- (5.36,2.57) -- cycle    ;
\draw    (51.61,9710.32) -- (51.61,9698.8) ;
\draw [shift={(51.61,9695.8)}, rotate = 90] [fill={rgb, 255:red, 0; green, 0; blue, 0 }  ][line width=0.08]  [draw opacity=0] (5.36,-2.57) -- (0,0) -- (5.36,2.57) -- cycle    ;

\draw    (129.5,9649.67) .. controls (140.48,9635.48) and (150.2,9640.98) .. (157.91,9647.76) ;
\draw [shift={(160,9649.67)}, rotate = 222.84] [fill={rgb, 255:red, 0; green, 0; blue, 0 }  ][line width=0.08]  [draw opacity=0] (10.72,-5.15) -- (0,0) -- (10.72,5.15) -- (7.12,0) -- cycle    ;

\draw   (117,9653.17) -- (126.5,9653.17) -- (126.5,9671.67) -- (117,9671.67) -- cycle ;

\draw  [color={rgb, 255:red, 0; green, 0; blue, 0 }  ,draw opacity=1 ] (164,9666.67) .. controls (164,9671.34) and (166.33,9673.67) .. (171,9673.67) -- (184,9673.67) .. controls (190.67,9673.67) and (194,9676) .. (194,9680.67) .. controls (194,9676) and (197.33,9673.67) .. (204,9673.67)(201,9673.67) -- (217,9673.67) .. controls (221.67,9673.67) and (224,9671.34) .. (224,9666.67) ;

\draw (210,9593) node [anchor=north west][inner sep=0.75pt]  [color={rgb, 255:red, 74; green, 144; blue, 226 }  ,opacity=1 ]  {$\delta _{v}$};
\draw (185,9630) node [anchor=north west][inner sep=0.75pt]  [color={rgb, 255:red, 0; green, 0; blue, 0 }  ,opacity=1 ]  {$w_{s}$};
\draw (185,9720) node [anchor=north west][inner sep=0.75pt]  [color={rgb, 255:red, 208; green, 2; blue, 27 }  ,opacity=1 ]  {$\delta _{th}$};
\draw (64.95,9705) node [anchor=north west][inner sep=0.75pt]  [font=\scriptsize]  {$\mathbf{e_{r}}$};
\draw (39,9686) node [anchor=north west][inner sep=0.75pt]  [font=\scriptsize]  {$\mathbf{e_{z}}$};
\draw (185,9685) node [anchor=north west][inner sep=0.75pt]  [color={rgb, 255:red, 0; green, 0; blue, 0 }  ,opacity=1 ]  {$\delta T_{s}$};

\end{tikzpicture}

\caption{Sketch of the side boundary layers when $Pr<1$.}
    \label{BL}
\end{figure}

We focus on the side wall considering the dimensionless steady axisymmetrical governing equations using cylindrical coordinates
\begin{equation}
u\frac{\partial u}{\partial r}+ w\frac{\partial u}{\partial z}=-Pr\frac{\partial P}{\partial r} +Pr~\nabla^2 u \ , 
\label{hormom}
\end{equation}
\begin{equation}
u\frac{\partial w}{\partial r} + w\frac{\partial w}{\partial z}=-Pr\frac{\partial P}{\partial z} + Ra_{\phi}PrT+ Pr\nabla^2 w \ , 
\label{vertmom}
\end{equation}
\begin{equation}
\frac{1}{r}\frac{\partial ru}{\partial r}+\frac{\partial w}{\partial z}=0 \ ,
\label{dimm}
\end{equation}
\begin{equation}
u\frac{\partial T}{\partial r} + w\frac{\partial T}{\partial z} = \nabla^2 T \ .
\label{dimnrj}
\end{equation}
Due to the low Prandtl regime, the viscous boundary layer is nested within the thermal one.
A diagram of this configuration is shown in Figure \ref{BL}.
The behavior of the vertical ($w_s$) and radial ($u_s$) velocities at the edge of the viscous boundary layer is derived from the conservation of mass \eqref{dimm} and energy \eqref{dimnrj}.
We assume that the typical scale of variation in the radial direction is the dimensionless thickness of the side boundary layer ($\partial/\partial_r \sim 1/\delta$). This thickness may either represent the thermal boundary layer ($\delta_{th}$) or the viscous boundary layer ($\delta_v$), depending on whether the radial variation under consideration pertains to temperature or velocity.
Moreover, we assume that the scale of variation in height is the dimensionless height equal to $1$ ($\partial/\partial_z \sim 1$). Then mass conservation \eqref{dimm} leads to 
\begin{equation}
\frac{u_s}{\delta_v}\sim w_s.
\label{reducem}
\end{equation}

On the left-hand side of  \eqref{dimnrj}, both advection terms, $u\partial_r T$ and $w\partial_z T$, scale as $w_s T$, due to mass conservation \eqref{reducem}. Because radial variations are much larger than height variations near the lateral boundaries, we assume that the dominant term in the Laplacian operator scales with the dimensionless thickness of the side boundary layer squared ($\nabla^2 \sim 1/\delta^2$). Therefore \eqref{dimnrj} leads to
\begin{equation}
\label{4}
w_s\sim 1/\delta_{th}^2.
\end{equation}
Let us now approximate the temperature variation across the thermal boundary layer ($\delta T_s$) by the average temperature of the system $\left<T\right>$ (recall that the dimensionless temperature on the side wall is zero).
Within the thermal boundary layer, we expect a force balance between the inertia term and the buoyancy term, so that the vertical momentum balance~\eqref{vertmom} reduces to 
\begin{equation}
u \pdv{w}{r} + w \pdv{w}{z} \simeq RaPrT.
\label{reducevertmom}
\end{equation}
On the left-hand side of  \eqref{reducevertmom}, each advection term, $u\partial_r w$ and $w\partial_z w$, scales like $w_s^2$, due to mass conservation \eqref{reducem}. Then \eqref{reducevertmom} reduces to $w_s^2 \sim Ra_{\phi}Pr\left<T\right>$ or equivalently, using \eqref{4},
\begin{equation}
\delta_{th}\sim\left(Ra_{\phi}Pr\left<T\right>\right)^{-1/4} \ .
\label{dlt}
\end{equation}
Finally, to link the heat flux applied on the bottom surface to the thermal characteristics of the side, a global flux balance is required.
Integrating \eqref{dimnrj} over the volume at steady state leads to:
\begin{equation}
\pi\Gamma^2\left(1-R_F\right)=2\pi\Gamma\frac{\left<T\right>}{\delta_{th}} \ ,
\label{bilan}
\end{equation}
where the left-hand side corresponds to the power mismatch between the lower and upper boundaries, which is balanced by the conducting flux across the thermal boundary layer on the right-hand side.
Thus, the averaged temperature $\left<T\right>$ is proportional to the aspect ratio and to the thickness of the thermal boundary layer 
\begin{equation}
\begin{aligned}
\left<T\right>\sim\delta_{th}(1-R_F)\Gamma \ .
\end{aligned}
\label{nu}
\end{equation}
Combining equations \eqref{dlt} and \eqref{nu}, the mean temperature can therefore be expressed in terms of the control parameters through the following relationship
\begin{equation}
\left<T\right>\sim Ra_{\phi}^{-1/5}~\Gamma^{4/5}~\left(1-R_F\right)^{4/5} Pr^{-1/5} \ .
\end{equation}

These simple dimensional arguments allow us to recover the scaling \eqref{scal} obtained via numerical simulations.
In addition, this reveals the dependencies on $R_F$ and $Pr$, which we did not observe since both these parameters have been fixed for now.
Notice that assuming that the average temperature is representative of the temperature difference across the radial thermal boundary layer is presumably only valid when the Rayleigh-Roberts number is large enough to mix efficiently the bulk of the convective system. In order to compare our results with existing literature, it is standard to use the Nusselt notion (the ratio of convective to diffusive heat flux). However, this measurement is only meaningful when the heat transfer can be unambiguously defined, that is, when the average isothermal surfaces are parallel. Or in other words, when temperature varies only along one dimension in the system.
In our system, there is no specific direction for heat transfer except in the asymptotic regime of high or low flux ratio. In these scenarios, it is conceivable to determine the Nusselt value, which indicates the main heat flux direction (vertical for $R_F \rightarrow 1$ and horizontal for $R_F=0$).

In the low flux ratio regime, the heat flux mainly goes in the horizontal direction, therefore we define the Nusselt by the relationship : 
\begin{equation}
\phi_{\textrm{in}} \equiv \frac{k}{R}\left(\left<\theta\right>-\theta_0\right)~Nu \ .
\label{nu_link}
\end{equation}
With the choice made for the non dimensional temperature, the Nusselt reads as the inverse of the mean temperature, hence
\begin{equation}
Nu~\sim~ Ra_{\phi}^{1/5}~\Gamma^{1/5}~\left(1-R_F\right)^{-4/5} Pr^{1/5}.
\end{equation}
Although we find a $1/5$ exponent for the Rayleigh dependency, reminiscent of the horizontal convection scaling of the \cite{ROSSBY} regime, it is important to note that the force balance in the boundary layer is different.
In \cite{ROSSBY}, buoyancy and viscosity at the bottom both play a role, while in our case, the balance is between inertia and buoyancy on the vertical side boundary.
The equivalent Rayleigh exponent based on a temperature scale (instead of a flux scale as done here) is $1/4$ and corresponds to the scaling of vertical convection identified by \cite{Batchelor}.
Furthermore, in a 2D rectangular system with identical thermal boundary conditions to the present case (except at the top boundary where $R_F$ was set to $0$), \cite{GANZAROLLI} identified a Nusselt scaling law. We obtain identical exponents for the Rayleigh-Roberts number and the aspect ratio as those found by \cite{GANZAROLLI}. Both exhibited $1/5$ exponents. It is worth mentioning that the validity of the Rayleigh scaling was recently shown to be limited to laminar boundary layers by \cite{shishkina}, so that a different scaling is expected at even large Rayleigh-Roberts numbers (we considered $Ra_{\phi}\leq10^9$). Previous works \citep{GEORGE} also pointed out this possibility. Finally, our results show the same $1/5$ Prandtl exponent dependence predicted by \cite{shishkina} and recently emphasized by \cite{zwirner}. However, we recall here that we have not checked this Prandtl scaling with our numerical simulations, which were all performed at $Pr=0.1$.

To validate the underlying physical approximations required to derive our scaling law, we now examine the secondary variables, specifically the thickness of the thermal boundary layer and the vertical velocity.
Thanks to the relations \eqref{4} and \eqref{nu}, the scaling laws for those quantities can be written as
\begin{equation}
\begin{aligned}
\delta_{th}&\sim Ra_{\phi}^{-1/5}~\Gamma^{-1/5}~\left(1-R_F\right)^{-1/5} Pr^{-1/5} \ ,\\
w_s&\sim Ra_{\phi}^{2/5}~\Gamma^{2/5}~\left(1-R_F\right)^{2/5} Pr^{2/5} \ .
\end{aligned}
\label{DV}
\end{equation}
\begin{figure}
   \centering
    \includegraphics[scale=0.32]{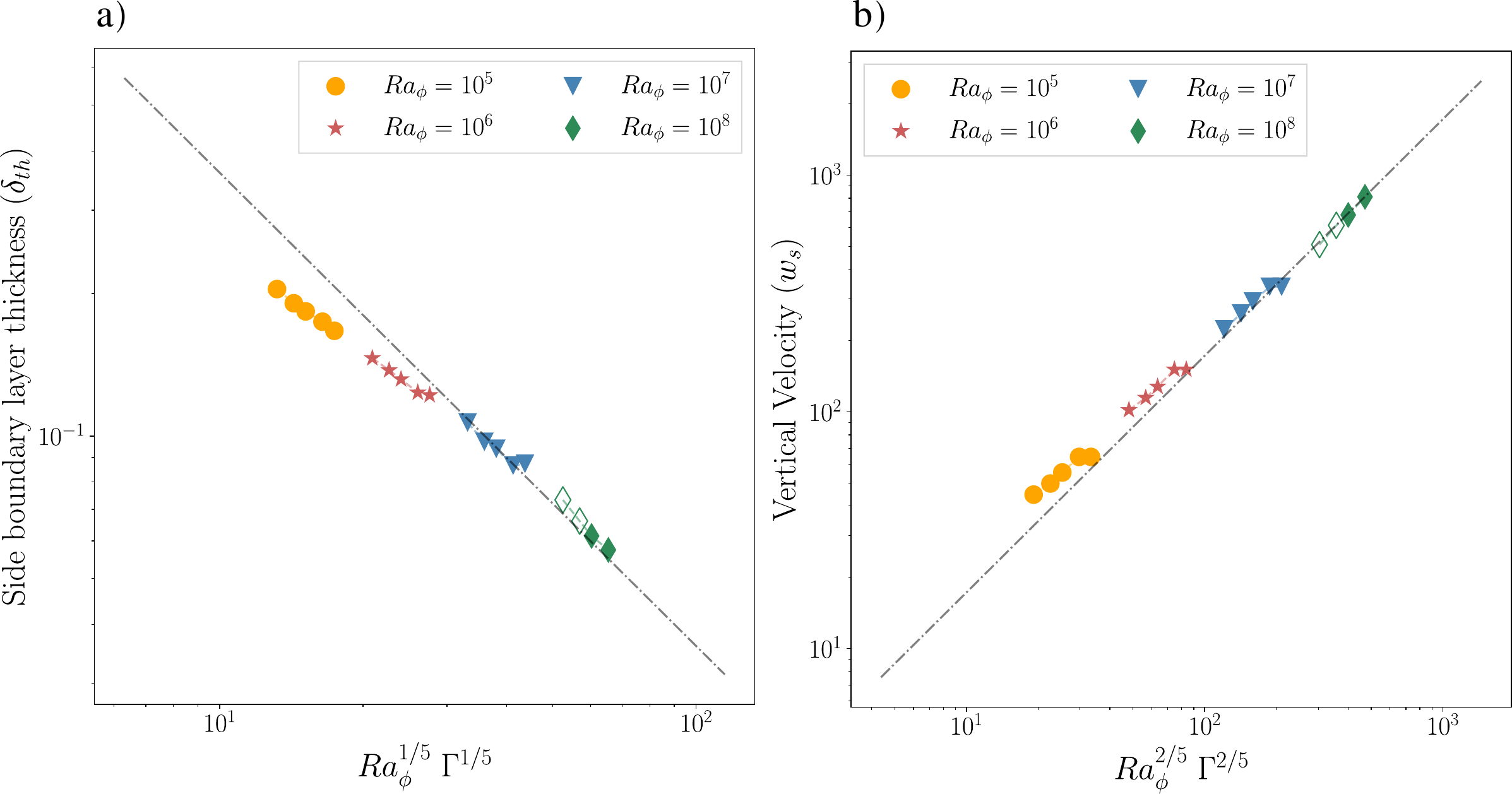}
   \caption{Log-log plot of a) the size of the thermal boundary layer and b) the absolute value of the vertical peak velocity near the edge (defined as the first minimum of the negative vertical velocity at mid-height, starting from the boundary and moving toward the center of the domain) vs. the determined scaling laws \ref{DV} for different $Ra_{\phi}$ and aspect ratios. The dash-dotted lines correspond to the scaling laws and for all the simulations. Empty/full symbols indicate respectively filtered/DNS simulations. The input parameters were $Pr=0.1$, $R_F=0.1$.}
   \label{w_d}
\end{figure}

The thickness of the thermal boundary layer is estimated by determining the radius at which $90\%$ of the maximum temperature near the edge is reached, while the vertical velocity is based on seeking the minimum vertical velocity near the edge. More precisely, we first compute the vertical velocity at mid-height of the domain to avoid the influence of the top and bottom boundaries, and we average in time and in the azimuthal direction. We then search for the first peak of negative vertical velocity, starting from the boundary and moving toward the center of the domain.
We conducted these measurements for aspect ratios of $4\leq\Gamma\leq16$ and for $10^5\leq Ra_{\phi}\leq10^8$.
Results shown in Figure \ref{w_d} are in excellent agreement with the scaling laws \eqref{DV}, hence further validating our approach.

\subsection{High flux ratio regime\label{highrf}}

We now consider the other limiting case of a flux ratio close to unity. We follow the same approach as in the previous section and we start with scalings obtained from numerical simulations followed by a theoretical explanation.

\subsubsection{Scaling for $R_F=0.9$}

\begin{figure}
   \centering
    \includegraphics[scale=0.32]{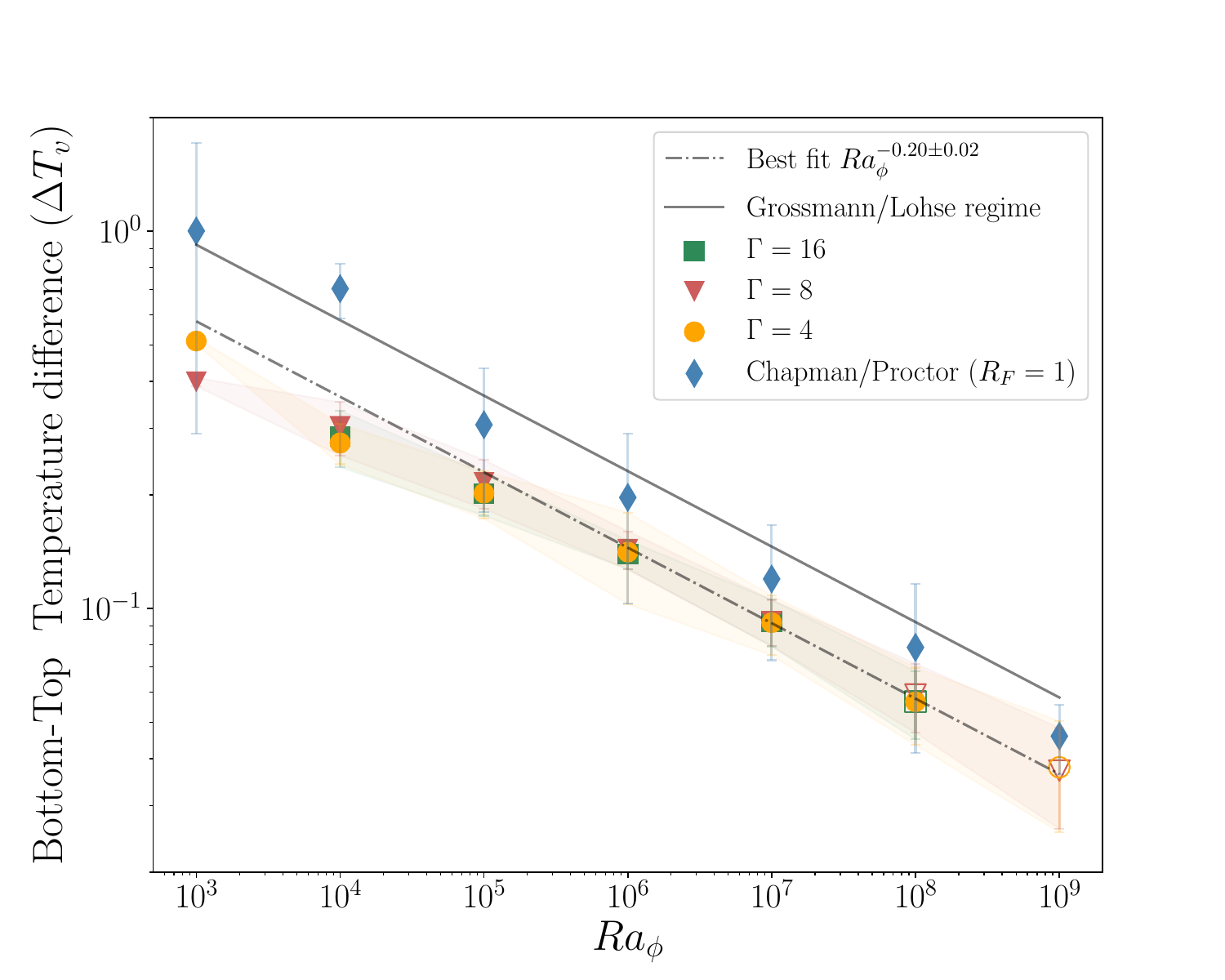}
   \caption{Log-log plot of the bottom-top temperature difference for different $\Gamma$ values. The error bars
are computed by taking 3 times the standard deviation of the bottom-top temperature difference time series
at the statistically stationary state. The dash-dotted line corresponds to the best fit scaling law $Ra_{\phi}^{-1/5}$, the blue diamond ($\Diamond$) plot shows the horizontally homogeneous \cite{cp} configuration ($R_F=1$ in a doubly-periodic Cartesian box) and the continuous black line corresponds to the \cite{GL} $I_l$ regime written in terms of flux Rayleigh number $Ra_{\phi}$. Empty/full symbols indicate respectively filtered/DNS simulations. The input parameters are $Pr=0.1$, $R_F=0.9$.
   \label{figure_RF09}}
\end{figure}

We now fix $R_F=0.9$, meaning that $90\%$ of the heating power goes out through the top surface.
Similarly to the precedent section, we seek power laws systematically varying $Ra_{\phi}$ and $\Gamma$.
However, in this second regime closer to the classical Rayleigh-Bénard configuration, heat transfers are mainly along the vertical direction.
We therefore focus on the mean temperature difference between the top and the bottom surface, denoted $\Delta T_v$ and computed as
\begin{equation}
\Delta T_v=\left<T(t,r,\varphi,z=0)-T(t,r,\varphi,z=1)\right>_{r\varphi} \ ,
\end{equation}
where the average corresponds to a temporal and surfacic average along radius and azimuthal angle.
As we will see below, this averaged quantity is more relevant than the mean temperature of the system which was more representative of the radial temperature differences between the bulk and the side boundary when $R_F=0.1$.
The evolution of the mean temperature difference with $Ra_{\phi}$ and $\Gamma$ is plotted in Figure \ref{figure_RF09}.
We observe that, as $Ra_{\phi}$ increases, $\Delta T_v$ decreases, suggesting that the system becomes more homogeneous vertically.
The decrease in the top-bottom temperature difference appears to be independent of the aspect ratio, which suggests a local mechanism.
Note also that this vertical temperature difference is the same at different locations, as will be further discussed below in section \ref{profiles}. 
When $Ra_{\phi}$ is larger than $10^5$, a power law scaling emerges with an exponent $Ra_{\phi}^{-0.20\pm 0.02}$ independently of $\Gamma$.
As we will show below, the closest relevant scaling is given by
\begin{equation}
\label{scalrf09}
    \Delta T_v\sim Ra_{\phi}^{-1/5} \ .
\end{equation}

\subsubsection{Theoritical analysis for $R_F=0.9$}

In this configuration close to Rayleigh-B\'enard convection, one might initially assume that the system is controlled by the heat flux across a thin thermal boundary layer \citep{malkus}. Let us define the Rayleigh number based on the temperature difference $Ra_{\Delta T_v}=(\beta g \Delta T_v H^3)/(\nu\kappa)$, which is an output parameter in our case since the temperature difference is not known a priori. Note that in the asymptotic limit where $R_F$ tends to one,  $Ra_{\Delta T_v}$ and $Ra_{\phi}$ are getting proportional to each other with the Nusselt number (as defined in Rayleigh-B\'enard convection) being the proportionality coefficient \citep{cioni_1997}.
The classical scaling $\Delta T_v\sim Ra_{\Delta T_v}^{-1/3}$ would equivalently give $Ra_\phi^{-1/4}$, which is not compatible with our result (\ref{scalrf09}).
 Our regime is closer to the regime $I_l$ predicted by \cite{GL} in which an energetic approach is used to estimate the viscous and thermal dissipation rates for the $Pr<1$ case.
In the $I_l$ regime, dissipation rates are dominated by their boundary layer contributions, therefore it is obtained at relatively low $Ra_{\Delta T_v}$ i.e, when the turbulence is sufficiently underdeveloped, so that the dissipation is mostly concentrated within the boundary layers.

Considering a homogeneous bulk (reached at sufficiently high Rayleigh-Roberts numbers) and thanks to the $I_l$ regime of \cite{GL}, the Nusselt scaling $Nu\sim\delta_b^{-1}\sim Ra_{\Delta T_v}^{1/4}$, where $\delta_b$ is the thickness of the bottom thermal boundary layer, leads to  $\Delta T_v \sim Ra_{\phi}^{-1/5}$ consistent with \eqref{scalrf09}.
It is noteworthy that analogy with the $I_l$ regime only makes sense if one assumes an equivalence between the upper and lower thermal boundary layers, which is not necessarily the case here since the velocity boundary conditions are mixed (free slip at the top, no slip at the bottom).
In addition, all the results of the study of \cite{GL} are obtained with an imposed temperature difference contrary to the imposed heat flux here.

When $R_F$ tends to $1$, our configuration is close to the one studied by \cite{cp}, where a constant flux is imposed at the bottom surface and goes out by the top surface ($R_F=1$) in a horizontally infinite domain ($\Gamma = \infty$).
This limiting case can be explored with a Cartesian box of size $2\times2\times 1$ periodic in both horizontal directions, with the same imposed flux at the top and the bottom and with no-slip and free-slip condition respectively at the bottom and top.
Figure \ref{figure_RF09} shows $\Delta T_v$ measured after the statistically-stationary state is obtained.
We also plot the temperature difference $\Delta T_v$ predicted by the $I_l$ regime scaling, keeping the prefactor determined for rigid boudaries and imposed temperatures \citep{GL}.
We find good agreement between the local Cartesian setup and the $I_l$ regime, with the same Rayleigh exponent being found.
A slight difference in the prefactor is nevertheless noted. Modifying the thermal and velocity boundary conditions did not have a significant impact, with only a minor change in the prefactor.
Our $R_F=0.9$ case is also found to be in good agreement with the $I_l$ regime, but with a small difference.
$10\%$ of the flux exits through the side, leading to a disruption in the vertical temperature gradient and therefore a difference in the temperature difference from bottom to top caused by the baroclinic flow.

\begin{figure}
   \centering
    \includegraphics[scale=0.32]{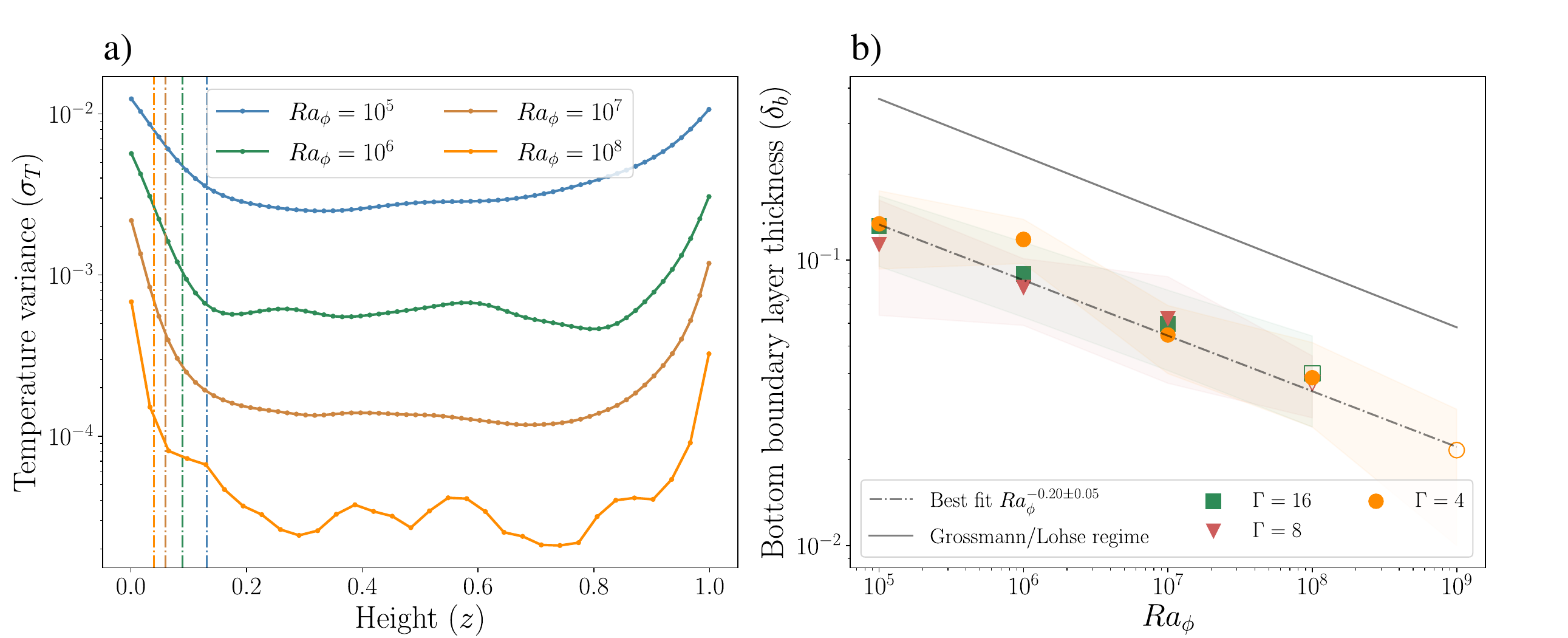}
   \caption{a) Plot of the temperature variance with the altitude $z$ for different $Ra_{\phi}$ at $\Gamma=16$, $Pr=0.1$ and $R_F=0.9$. The irregular profile for $Ra_{\phi}=10^8$ is due to a lack of statistical samples for this very costly computation. The vertical dotted lines indicate the boundary layer altitudes based on the $90\%$  decrease of the temperature variance near the bottom. b) Log-log plot of the bottom boundary layer with $Ra_{\phi}$ for different $\Gamma$. The continuous black line represents the \cite{GL} scaling corresponding to the $I_l$ regime. Empty/full symbols indicate respectively filtered/DNS simulations. The input parameters were $Pr=0.1$, $R_F=0.9$.}
   \label{figure_bl09}
\end{figure}
Another way to verify the relevance of the $I_l$ regime is to check the bottom thermal boundary layer behaviour.
The scaling law predicts that $\delta_{b} \sim  Ra_{\phi}^{-1/5}$.
A measure of the bottom thermal boundary layers has been done based on the temperature variance computed as
\begin{equation}
    \sigma_T=\Big< \big(T(t,r=\Gamma/2,\varphi,z)-\left<T(t,r=\Gamma/2,\varphi,z)\right>_{\varphi}\big)^2\Big>_{\varphi}^{1/2} \ .
\end{equation}
We focus here on the temperature at $r=\Gamma/2$ to avoid the side and center areas which are more significantly affected by the overall baroclinic circulation driven at the sidewall.
In Figure \ref{figure_bl09}(a), $\sigma_T$ as a function of the altitude $z$ is plotted for different $Ra_{\phi}$ and for $\Gamma=16$.
Near the top and bottom boundaries, the temperature variance is rapidly varying, which indicates the existence of boundary layers.
In the bulk $0.2<z<0.8$, the system is more homogeneous especially when $Ra_{\phi}$ is high.
Vertical dotted lines indicate the thickness of the bottom boundary layer $\delta_b$.
It is estimated as the position $z$ for which the temperature variance has decreased by $90\%$ compared to its value at the boundary. We used the temperature variance rather than the vertical temperature profile because the recirculation flow (induced by the cold side) involves a heat transport mechanism similar to the horizontal convection \citep{mullarney_2004} altering the thermal boundary layer structure.
In Figure \ref{figure_bl09}(b), the corresponding boundary layer thickness is plotted as a function of $Ra_{\phi}$ for several $\Gamma$ at $R_F=0.9$.
It is independent of $\Gamma$ and scales like $\delta_b \sim Ra_{\phi}^{-0.20\pm 0.05}$ (determined by the least square method).
The $I_l$ regime law \citep{GL} is also plotted and is consistent with our measurements at $R_F=0.9$ in terms of exponent, with again an offset on the prefactor presumably due to residual effects of the large-scale circulation driven at the side wall.

\subsection{Intermediate regimes}

In sections~\ref{lowrf} and \ref{highrf}, we have identified two different temperature averages which seem to characterise the system behavior in the low/high flux ratio regimes.
We now investigate the case $R_F=0.5$ where the heat flux equally goes out through the top and the side.
As before, we seek power laws systematically varying $Ra_{\phi}$ and $\Gamma$.
In the precedent limiting regimes, heat transfers were dominantly radial or vertical, but the outgoing flux is now evenly distributed between the top and the side boundaries, so that a mixed state is expected.
Thus, we now carry out an analysis using both the mean temperature $\left<T\right>$ and the top/bottom temperature difference $\Delta T_v$.

\begin{figure}
   \centering
    \includegraphics[scale=0.28]{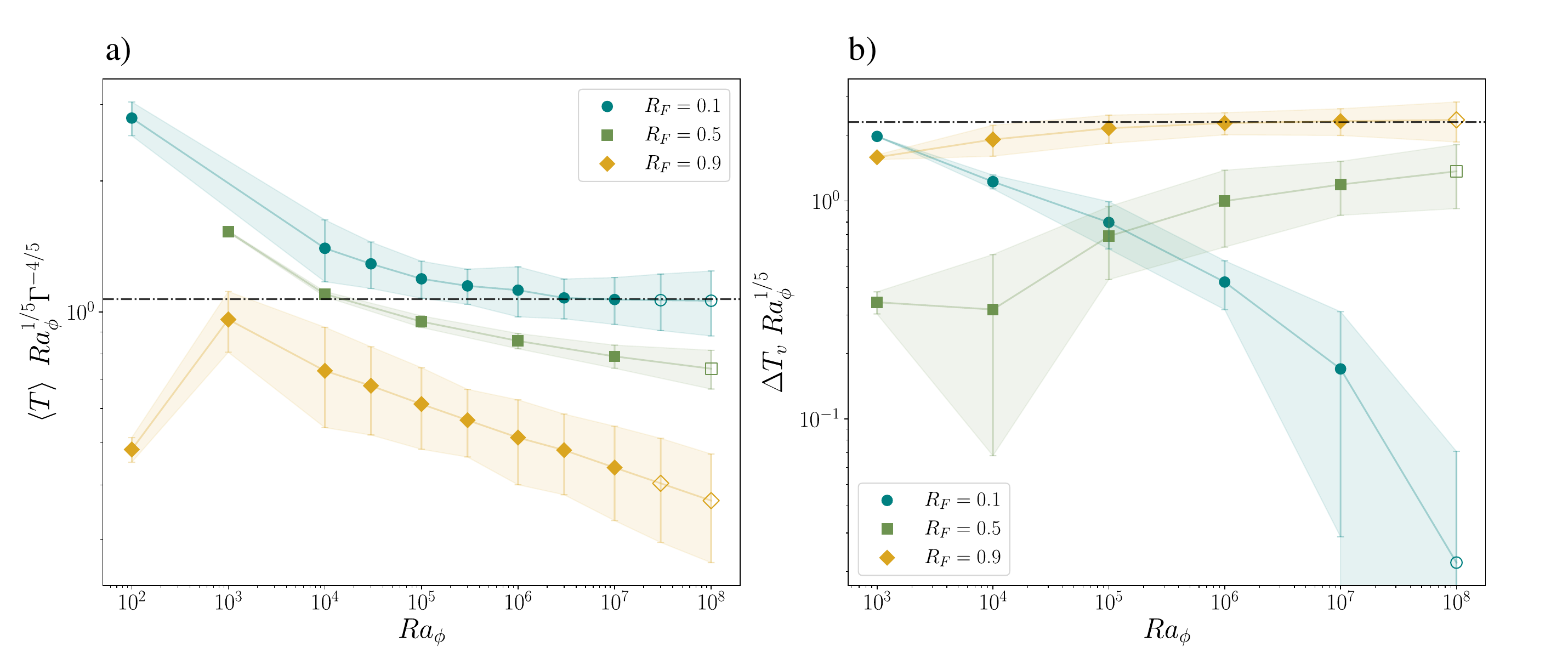}
   \caption{a) Compensated mean temperature, b) Compensated bottom-top temperature difference as a function of $Ra_{\phi}$ for $R_F=0.1~(\circ)$, $R_F=0.5~(\square)$, $R_F=0.9~(\Diamond)$. The error bars are computed by taking 3 times the standard deviation of the time series at the statistically stationary state multiplied by
   $Ra_{\phi}^{1/5}\Gamma^{-4/5}$ and $Ra_{\phi}^{1/5}$ respectively. Empty/full symbols indicate respectively
filtered/DNS simulations. The input parameters are $Pr=0.1$ and $\Gamma=8$.}
   \label{fig_rf05}
\end{figure}

\begin{figure}
   \centering
    \includegraphics[scale=0.3]{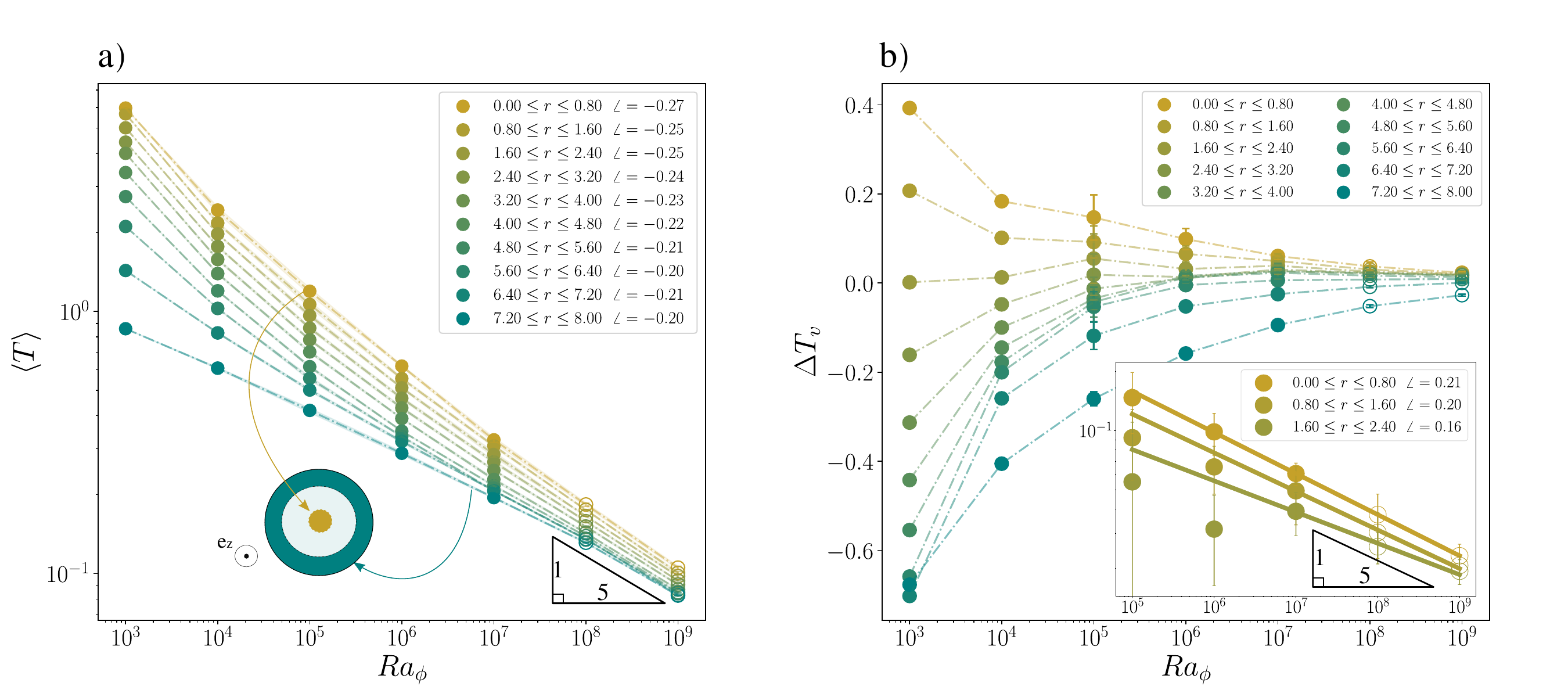}
   \caption{a) Mean temperature  $\left<T\right>$ and b) bottom-top temperature difference $\Delta T_v$, defined on successive rings with $Ra_{\phi}$ . The step radius is $0.8$, starting from $0$ to $0.8$ and ending from $7.2$ to 8. For a), the power law exponent($\angle$) is measured for each ring. The inset in b) shows a log-log plot of $\Delta T_v$ and associated power law exponents for the most inner rings. Empty/full symbols indicate respectively filtered/DNS simulations. The input parameters are $Pr=0.1$, $\Gamma=8$ and $R_F=0.1$. }
   \label{Crown}
\end{figure}

Figure \ref{fig_rf05} shows plots of both the mean temperature $\left<T\right>$ and the bottom-top temperature difference $\Delta T_v$ compensated by the low/high flux ratio scaling law respectively, for a fixed value of $\Gamma=8$ and for the three flux ratios $R_F\in[0.1,0.5,0.9]$.
The best-fit power laws for $\left<T\right>$ and $\Delta T_v$ at the intermediate flux ratio of $R_F=0.5$ do not correspond to the scaling laws observed in previous regimes, exhibiting dependencies of $Ra_{\phi}^{-0.23\pm 0.02}$ and $Ra_{\phi}^{-0.13\pm 0.11}$ respectively. To clarify these observations, let us define the following averaging operator:

\begin{equation}
    \left<T\right>(r_1<r<r_2)=\frac{1}{\tau \pi(r_2^2-r_1^2)}\int_{t_0}^{t_0+\tau}\int_0^1\int_0^{2\pi}\int_{r_1}^{r_2}T~r\textrm{d}r\textrm{d}\varphi\textrm{d}z\textrm{d}t \ ,
\end{equation}
representing a temporal and volume average restricted to a particular radius range $r_1<r<r_2$.
A similar definition without vertical integration is used for the top-bottom temperature difference.
In Figure \ref{Crown}, we show such local averages for various increasing values of the limiting radii $r_1$ and $r_2$.
We fix $R_F=0.1$, $\Gamma=8$ and $10^3<Ra_{\phi}<10^9$.
The mean temperature exhibits different power laws with varying exponents in different radial regions.
Close to the side boundary ($7.2<r<\Gamma=8$ typically), the local average $\left<T\right>$ follows a $Ra_{\phi}^{-1/5}$ scaling, corresponding the low flux ratio regime as expected from the proximity with the sidewall circulation.
In the bulk however ($0<r<0.8$ typically), an exponent of $-0.27$ is obtained, reminiscent of the scaling observed for the high flux ratio regime.
This implies that even at a flux ratio $R_F=0.1$, the high flux ratio mechanism persists in some form close to the center.
$\Delta T_v(r_1<r<r_2)$ shown on Figure \ref{Crown}(b) also exhibits a regime transition.
For $r>4$, $\Delta T_v$ becomes negative at low $Ra_{\phi}$ and increasingly negative towards the edge, highlighting the sidewall impact on the dynamics.
In the centre region however ($0<r<0.8$), the power law for $\Delta T_v$ is consistent with the $I_l$ \cite{GL} regime, following a $Ra_{\phi}^{-1/5}$ dependency, as expected from the quasi-homogeneous behaviour of the bulk convection far from the sidewall.

The system is therefore inherently inhomogeneous with a permanent interaction between two limiting regimes:
one in the bulk defined by the $I_l$ \cite{GL} regime, and the other one near the edge defined by vertical natural convection. In the asymptotic regimes, the limiting mechanism (side and bottom/top boundary layers respectively for $R_F=0.1$ and $0.9$) controls the overall heat transfers and thermal structure. The variation of the flux ratio parameter involves a continuous transition with a gradual shift from the dominance of one regime to another. But for any flux ratio, signatures of both regimes can be seen in the radial profiles, as will be studied below. The regime interaction artifically arises from describing an inhomogeneous system using global variables ($\left<T\right>$, $\Delta T_v$). In fact, these two mechanisms operate simultaneously but in different locations. The high flux ratio regime is primarily located near the center of the domain, while the low flux ratio regime is predominantly located at the side. 
Our study hence shows that the global mean temperature values $\left<T\right>$ and $\Delta T_v$ are not adequate to fully characterise the system at any flux ratio due to the presence of two distinct mechanisms, and the radial inhomogeneities in the system statistics, as discussed in the next section.

\subsection{Radial inhomogeneities\label{profiles}}

\begin{figure}
   \centering
    \includegraphics[scale=0.4]{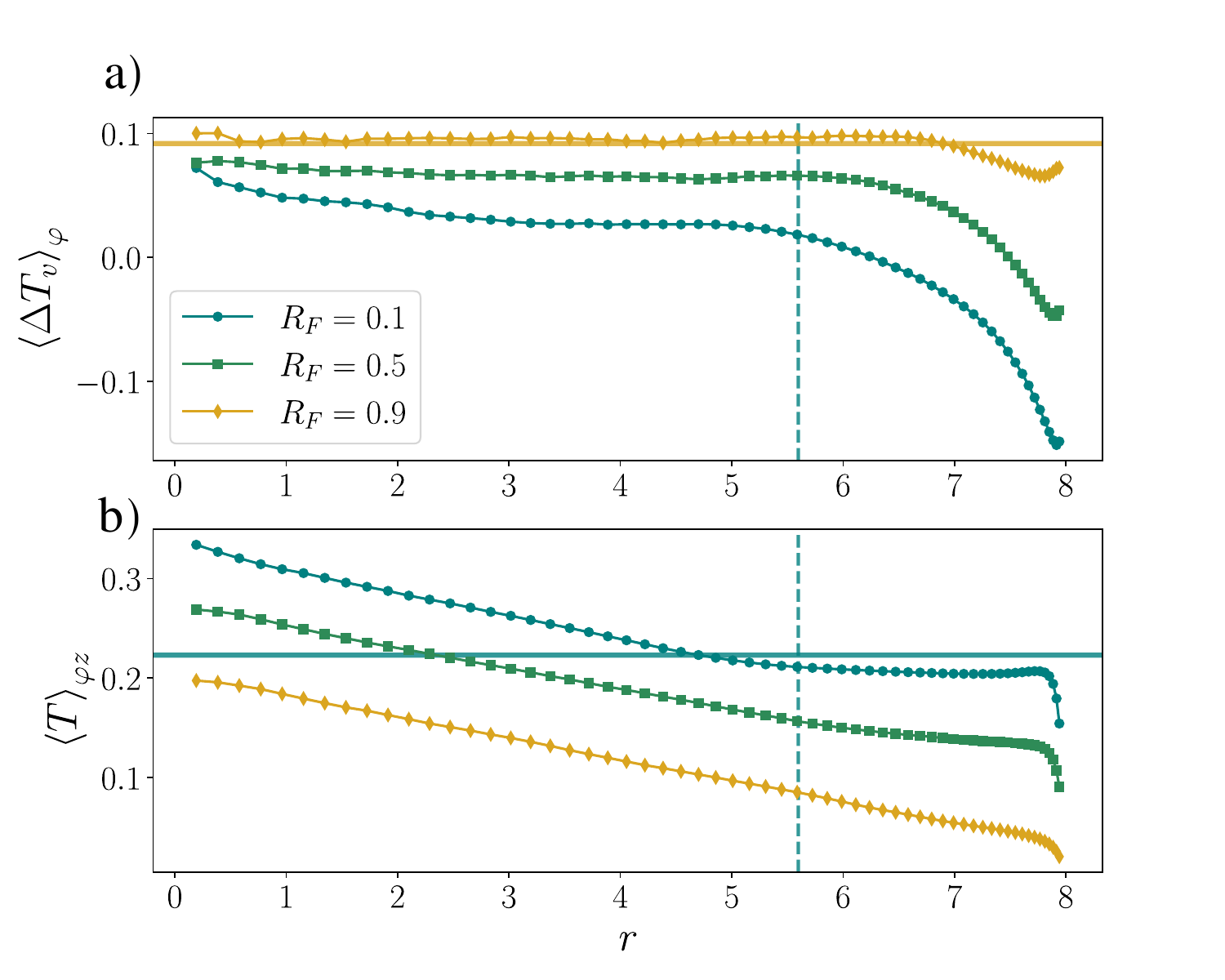}
   \caption{Radial profile of the bottom/top temperature difference a) and of the mean temperature b) for $R_F=0.1$ ($\square$), $R_F=0.5$, ($\Diamond$) and $R_F=0.9$ ($\circ$). The horizontal lines indicate the radial-mean value for $R_F=0.9$ in a) and $R_F=0.1$ in b), while the dash lines indicate the $r_p$ value for $R_F=0.1$. The input parameters are $Ra_{\phi}=10^7$, $\Gamma=8$, and $Pr=0.1$.}
   \label{Tprofil}
\end{figure}

In this section, we study the radial structure of the two temperature variables examined previously. 
In Figure \ref{Tprofil}, we present the spatio-temporal averages $\left<T\right>_{\varphi z}$ and $\left<\Delta T_v\right>_{\varphi}$ as a function of radius, for different flux ratio values with $Ra_{\phi}=10^8$ and $\Gamma=8$.
For $R_F=0.9$, $\Delta T_v$ is constant on a large part of the domain except near the sidewall where an inhomogeneity is clearly visible.
The lower the $R_F$ the larger this inhomogeneous domain, as illustrated in Figure~\ref{Tprofil}a). Indeed, as $R_F$ decreases, the sidewall circulation gets stronger and perturbs the bulk forced convection which tends to vertically homogenise the temperature. 
The radial evolution of the mean temperature (Figure~\ref{Tprofil}b) also reveals two distinct regions.
One region is located near the lateral sidewall and displays a nearly-constant temperature (excluding the thin boundary layer developing along the sidewall) in good agreement with the volume-averaged temperature at low flux ratio, shown as a thin horizontal line for $R_F=0.1$.
The second, inner region shows a linear increase in temperature towards the center.
An increase in the prescribed heat flux at the top affects the radial temperature profile.
Indeed, when $R_F=0.9$, the uniform temperature zone is getting pushed towards the sidewall, while the same temperature gradient is observed in the central region.
Regarding the intermediate case $R_F=0.5$, the two distinct areas previously identified can still be seen, but the plateau zone is smaller compared to the case $R_F=0.1$.
The size of this region is directly tied to the magnitude of the lateral heat flux and will be further studied in the next section by analysing the flow structure.

In view of these two zones clearly identified in the radial temperature profile, we suggest to approximate it (outside the outer thermal boundary layer) by considering the following three parameters model 
\begin{equation}
\left<T\right>_{\varphi z}(r)=\left\{\begin{array}{ll}
        G_{T}~r_p\left(1-\frac{r}{r_p}\right)+T_p & \mbox{for}~r\leq r_p\\
         T_p & \mbox{for}~r\geq r_p
    \end{array}
\right. \label{eq:1Dmodel}
\end{equation}
with the constant temperature near the sidewall $T_p$, the slope of the temperature profile in the linear region $G_T$ and the  transition radius $r_p$.
In the next sections, we focus on the determination of each of these parameters and discuss their physical origin.

\subsubsection{Radial temperature gradient}
\begin{figure}
   \centering
    \includegraphics[scale=0.3]{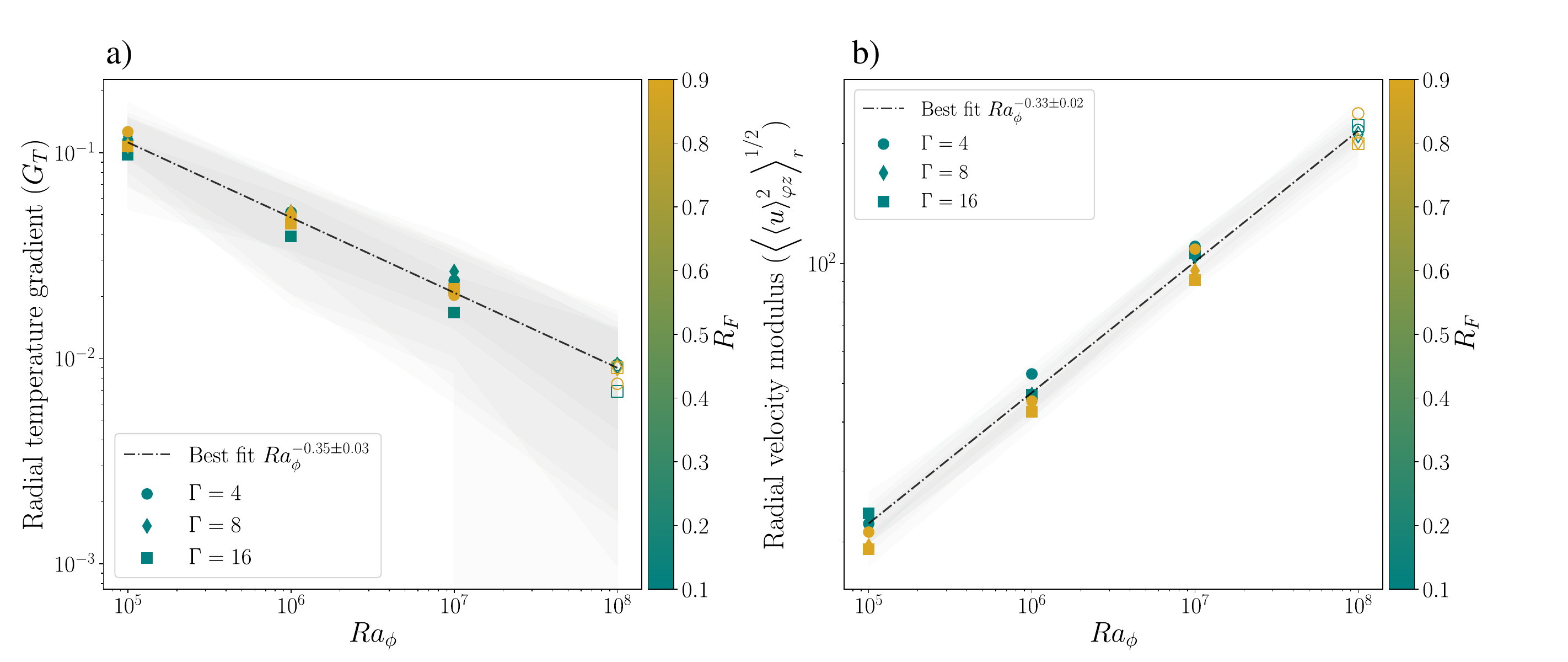}
   \caption{Log-log plot of the radial temperature gradient in a) and of the RMS radial velocity in b) as a function of $Ra_{\phi}$, with various values of $\Gamma$ shown with symbols and of $R_F$ shown with colors. The dash-dotted lines indicate the best fit power law. The colored areas are computed by taking the standard deviation of the radial temperature gradient and the square of the radial velocity in the bulk for a) and b) respectively. Empty/full
symbols indicate respectively filtered/DNS simulations and for all $Pr=0.1$.}
   \label{figure_grad_scal}
\end{figure}

First, let us compute the slope of the radial temperature profile $G_T$ as $\left |\left<\frac{\mathrm{d}\left<T\right>_{\varphi z}}{\mathrm{d}r}\right>_{r} \right |$ and the radial velocity modulus computed as $\sqrt{\left<\left<u\right>_{\varphi z}^2\right>_r}$ restricted to $r\leq5/8\Gamma$ and  $r\leq0.95\Gamma$ respectively for $R_F=0.1$ and $0.9$, $10^5<Ra_{\phi}<10^8$ and considering $\Gamma=4,8$ and $16$.
All data are plotted in Figure \ref{figure_grad_scal}.
Both the mean temperature gradient and the radial velocity do not significantly depend on the flux ratio, nor on the aspect ratio, and they seem to be anti-correlated with each other.
Indeed, they both follow a power law behaviour with $Ra_{\phi}$, with the same approximate exponent $1/3$ but positive for the radial velocity and negative for the temperature gradient.
This suggests that, to understand the physical origin of the thermal gradient, a closer look at the flow structure is necessary.

\begin{figure}
   \centering
    \includegraphics[scale=0.3]{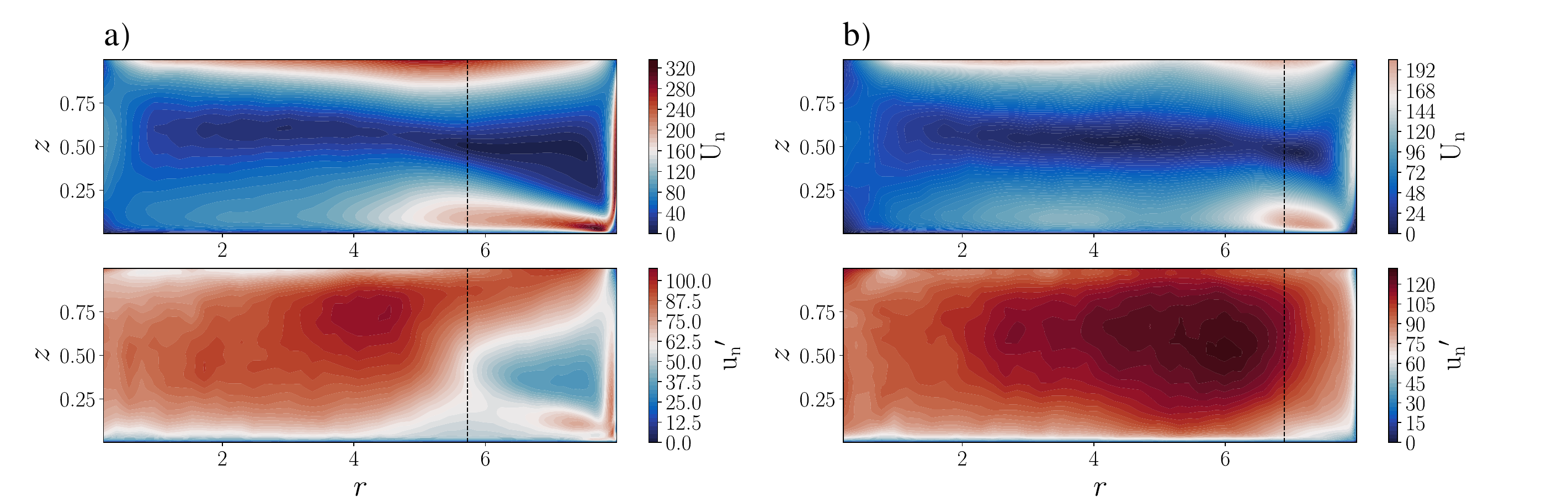}
   \caption{Map in the $r,z$ plane of the mean (top) and fluctuating (bottom) velocity norm for a) $R_F = 0.1$ and b) $R_F = 0.9$. The dashed lines indicate the $r_p$ value. The input parameters are $Pr = 0.1$, $\Gamma= 8$ and $Ra_{\phi}=10^7$.}
   \label{figTT}
\end{figure}
To do so, we consider the representative case $Ra_{\phi}=10^7$ and $\Gamma=8$.
We compute in a $r,z$ plane and for both regimes $R_F=0.1$ and $R_F=0.9$, the norm of the velocity field denoted $\mathrm{U_n}$, shown in the top panels of Figure \ref{figTT}, as well as the velocity fluctuations field denoted $\mathrm{u_n}^\prime$, shown at the bottom.
Those fields are computed as follows
\begin{equation}
\mathrm{U_n}(r,z)=\left[\left<u\right>_{\varphi}^2+\left<w\right>_{\varphi}^2\right]^{1/2},~~\mathrm{u_n}^\prime(r,z)=
\left[\left<\left(u-\left<u\right>_{\varphi}\right)^2\right>_{\varphi}+\left<\left(w-\left<w\right>_{\varphi}\right)^2\right>_{\varphi}\right]^{1/2} \ .
\end{equation}
In both regimes, a global circulation surrounds the whole domain, with intense mean velocities close to the boundaries.
In the bulk, fluctuations dominate the flow, at least for $0<r<5$ for $R_F=0.1$ and up to the side boundary layer for $R_F=0.9$. These regions correspond to the domain with a significant radial gradient of the mean temperature, shown in Figure~\ref{Tprofil}.
In order to understand the emergence of the radial temperature gradient, let us look at the heat transport equation.
We average it in time, in azimuth, but also over a particular thickness $h$.
This thickness corresponds to the altitude at which the vertical profile of the radial velocity $\left<u\right>_{\varphi r}$ changes sign.
We denote this averaging operation as $\left<\bullet\right>_{\varphi h}$. 
To clarify, we do not average over the whole depth because the terms related to radial advection would vanish owing to continuity.
Our averaging procedure allows to distinguish between the mean flow carrying hot fluid from the center to the edge of the domain in the upper region $z>h$ and cold fluid advected from the edge to the center in the bottom region $z<h$.

By decomposing the temperature and velocity fields between mean and fluctuating components, the heat equation can be written as
\begin{equation}
\frac{\partial T'}{\partial t} +(\bm{\mathcal{U}}+\bm{u'})\bm{\cdot\nabla }(\mathcal{T}+T')=\bm{\nabla}^2 (\mathcal{T}+T') \ ,
\end{equation}
where $\bm{\mathcal{U}}=(\mathrm{U},\mathrm{V},\mathrm{W})^T=\left<\bm{u}\right>_{\varphi h}$ and $\mathcal{T}=\left<T\right>_{\varphi h}$ represent respectively the velocity and temperature mean component and $\bm{u'}=(u',v',w')^T,~T' $ the fluctuating ones.\\

Thus, the heat equation reads
\begin{equation}
\mathrm{U}\frac{\mathrm{d}\mathcal{T}}{\mathrm{d}r}+\left<u'\frac{\partial  T'}{\partial r}\right>_{\varphi h} +\left<(\mathrm{W}+w')\frac{\partial  T'}{\partial z}\right>_{\varphi h}=\frac{1}{h} + \frac{1}{r}\frac{\mathrm{d}}{\mathrm{d} r}\left(r\frac{\mathrm{d}\mathcal{T}}{\mathrm{d} r}\right)+ \frac{1}{2\pi h}\int_{0}^{2\pi}\frac{\partial T'}{\partial z}_{|z=h}  \mathrm{d\varphi} \ .
\label{eq:meant}
\end{equation}

Neglecting the mean vertical transport (third term on the left-hand side), diffusive terms (second and third terms on the right-hand side) and the transport induced by fluctuations (second and fourth terms on the left-hand side), equation~\eqref{eq:meant} reduces to
\begin{equation}
\frac{\mathrm{d}\mathcal{T}}{\mathrm{d}r} \simeq \frac{1}{\mathrm{U}h} \ .
\label{eq_slope}
\end{equation}

The radial temperature gradient appears to be inversely proportional to the radial velocity.
Note that our approach here was guided and validated by empirical observations, and we did not perform any formal ordering between the various terms of (\ref{eq:meant}).
\begin{figure}
   \centering
    \includegraphics[scale=0.28]{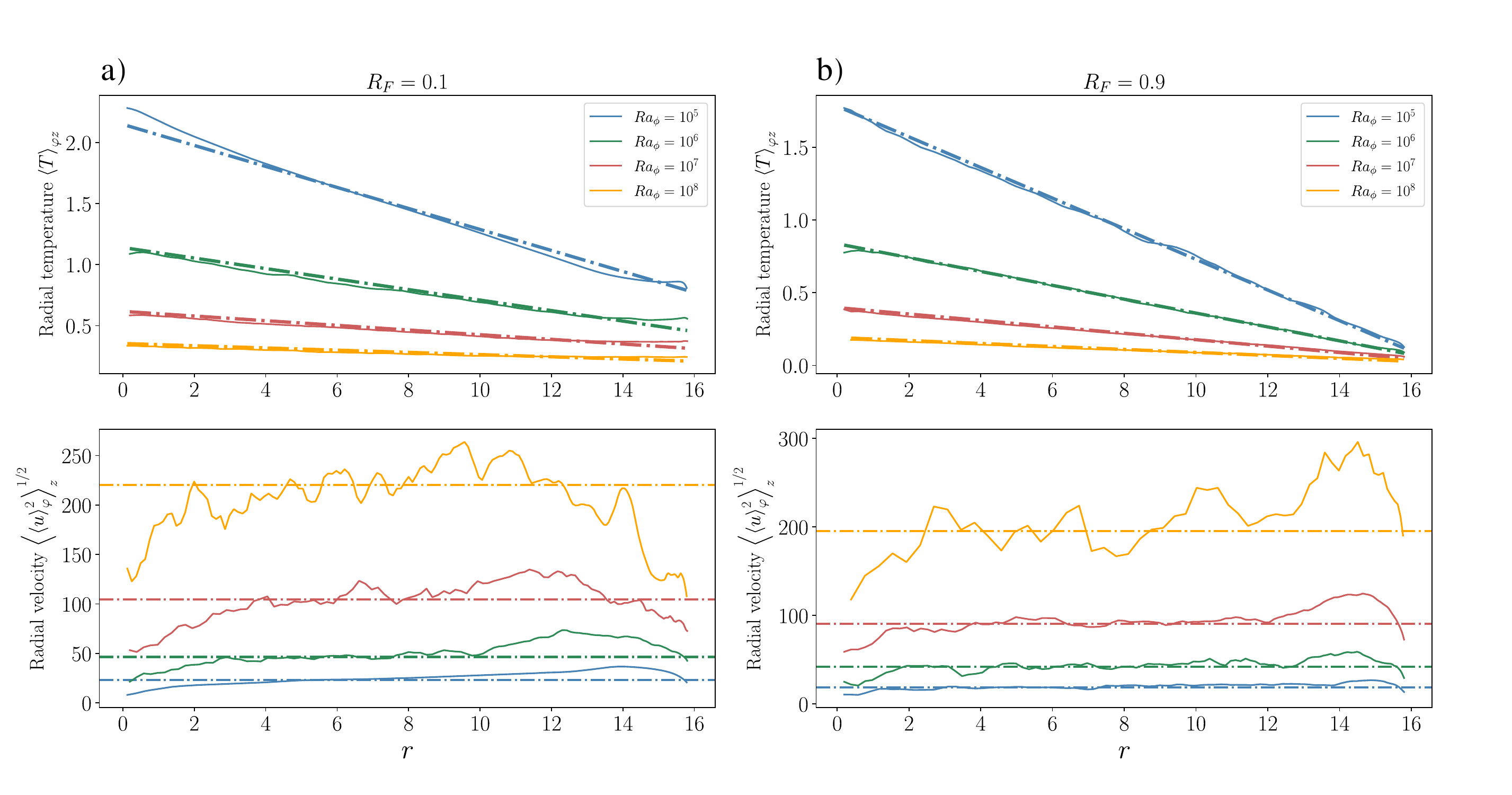}
   \caption{Radial profile of the mean temperature (top row) and of the radial velocity modulus (bottom row) for different $Ra_{\phi}$ and for $a)~R_F=0.1$ and $b)~R_F=0.9$. The other input parameters are $Pr=0.1$ and $\Gamma=16$. The dash-dotted lines correspond to the theoretical slope~\eqref{eq_slope} for the top row and to the mean value of the radial velocity modulus for the bottom row.}
   \label{figure_slope}
\end{figure}
In figure \ref{figure_slope}, we plot the radial temperature profiles and the radial velocity modulus profiles for both regimes $R_F=0.1$ and $R_F=0.9$, with $\Gamma=16$ and $Ra_{\phi}$ between $10^5$ and $10^8$. We also plot the theoretical radial temperature gradient estimated by (\ref{eq_slope}).
To do so, $\mathrm{U}$ was estimated as the mean value of the radial velocity modulus ($\left<|\left<u\right>_{\varphi}|\right>_z$) when $2<r<10$ and $2<r<12$ respectively for $R_F=0.1$ and $R_F=0.9$ i.e. when the velocity is relatively constant.
It can be see in figure \ref{figure_slope}, when $Ra_{\phi}$ increases, the values of the radial velocity modulus profile increases while the radial temperature gradient decreases.
Furthermore, we can see that our estimate~\eqref{eq_slope} matches very well to the radial temperature profiles observed in simulations.

This model derives from the assumption that the mean flow is responsible for the radial temperature gradient.
One could have argued that on the contrary, the radial temperature gradient creates the flow by a baroclinic torque: the temperature gradient would then be proportional (and not inversely proportional as we observe) to $U$.
Thus, as will be further detailed in the next sections, we conclude that the side cold temperature generates the mean flow which then builds up the radial temperature gradient.
Incidentally, the radial gradient scaling with $Ra_{\phi}$ shown in Figure \ref{figure_grad_scal},  $G_T\sim Ra_{\phi}^{-1/3}$, implies that it decreases faster with $Ra_{\phi}$ than the mean temperature of the system ($\left<T\right>\sim Ra_{\phi}^{-1/5}$, see \eqref{scal}).
This hints that the radial thermal gradient is a secondary aspect of the thermal structure.
At first order, the plateau scaling is the dominant factor, which explains why the scaling of the low flux ratio regime works well at high $Ra_{\phi}$ values for all $R_F$, at least in the vicinity of the sidewall.

\subsubsection{Temperature plateau value}
\begin{figure}
   \centering
    \includegraphics[scale=0.4]{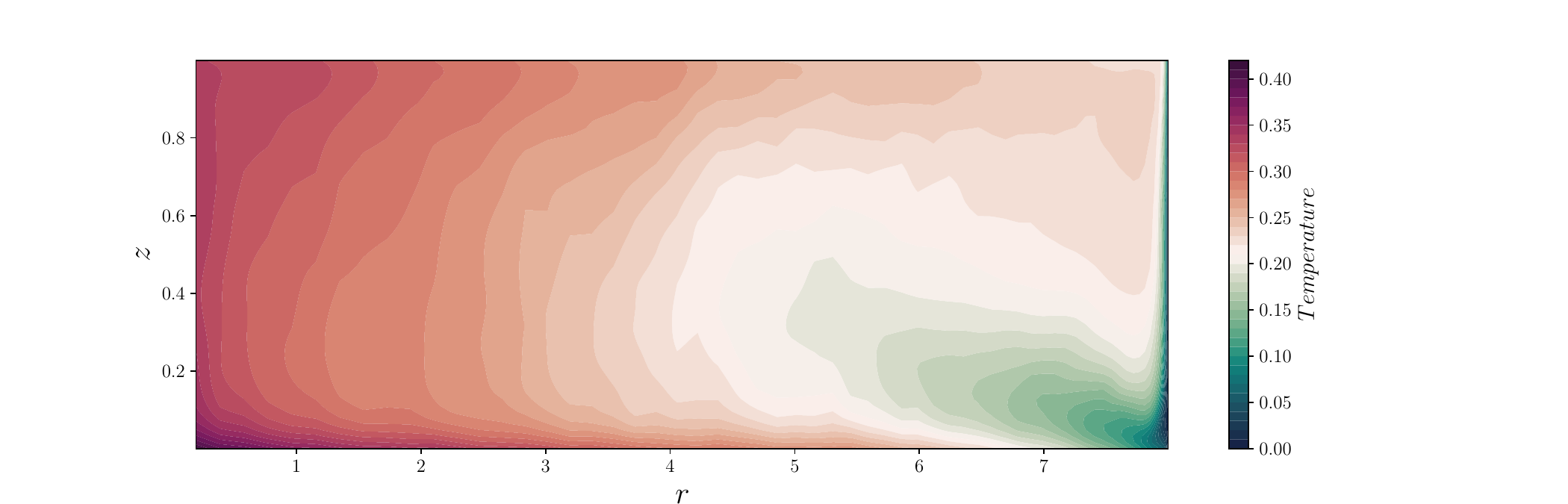}
   \caption{$r,z$ map of the temperature field averaged in time and azimuthal direction. The input parameters are $Ra_{\phi}=10^7$, $\Gamma=8$, $R_F=0.1$ and $Pr=0.1$.} 
   \label{figTmap}
\end{figure}

The physical reason behind the uniform radial temperature profile near the sidewall, particularly visible at low $R_F$, is not trivial.
Indeed, Figure \ref{figTmap} shows that the temperature map averaged over azimuth and time, but not over depth, is very heterogeneous along the vertical direction.
A cold zone is localised at the bottom close to the edge, in close connection with the localized, strong velocity zone seen in Figure \ref{figTT}(a). Indeed, close to the sidewall, there is a strong downward flow which then turns into a cold jet with a characteristic radial extent, seemingly corresponding to the plateau area on the temperature profile.
The larger $Ra_{\phi}$, the more extended this area.
Actually, as we shall see, when $Ra_{\phi}$ increases, the cold jet which drives the global recirculation gets more inertia and propagates more into the domain. Averaging over depth, the $T_p$ plateau value is well predicted by the equation \ref{nu} resulting from the analysis of the heat transfer thought the side boundary layer made in the first regime. We now focus on understanding the cold jet dynamics in order to estimate the $r_p$ parameter.

\subsubsection{Cold jet penetration length}

We assume that the cold jet is a turbulent, self-similar structure, where a given amount of momentum initially injected at the side is propagating in the radial direction and heated from below by the lower plate.
Following the seminal work of \cite{morton1956turbulent}, it is well known that the thickness of jets and thermal plumes increases linearly along their propagation direction due to the turbulent entrainment of the ambient fluid \cite[see also, e.g.,][]{ANR_list,turner_1986}.
\begin{figure}
   \centering
    \includegraphics[scale=0.35]{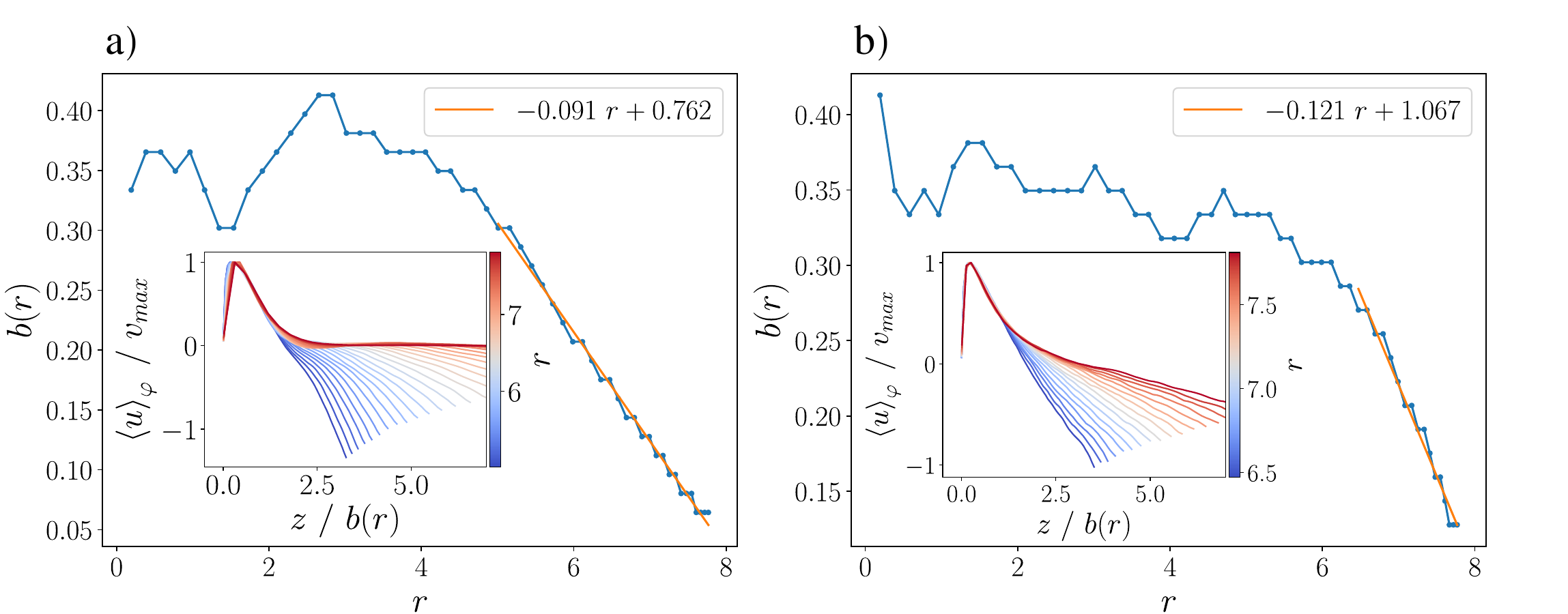}
   \caption{Plot of the cold plume thickness $b$ as a function of the radius. The blue points are the data, while the continuous line in orange is the best fit. The insets show the radial velocity normalised by the maximal velocity inside the jet, as a function of the height normalized by the jet thickness. Several radii indicated by the colorbar are shown going from near the edge (red) to the bulk (blue). $Ra_{\phi}=10^8$, $\Gamma=8$, $Pr=0.1$ and $R_F=0$ and $0.7$ respectively for $a)$ and $b)$. }
   \label{coldtonguefig}
\end{figure}

Our structure can be seen as a mixture between a jet and a plume, and so it is natural to fit its thickness by a linear growth as 
\begin{equation}
b = \alpha (\Gamma-r) + r_0,
\end{equation}
where $r_0$ is the radius close to the sidewall and $\alpha$ a constant known as the entrainment coefficient.
In figure \ref{coldtonguefig}, the thickness of the cold jet is plotted as a function of the radius for the case where $Ra_{\phi}=10^8$, $\Gamma=8$, and for $R_F=0$ and $0.7$.
We estimate the cold jet thickness at a given $r$ by seeking the depth where the ratio between the radial velocity (averaged in time and azimuthal direction) and the maximal value of this velocity located in the cold jet is equal to $0.5$.
We see the linear behaviour corresponding to the cold jet spreading starting near the edge and stopping where $r\approx 5/8 \Gamma$ for $R_F=0$ and $r\approx 0.8\Gamma$ for $R_F=0.7$.
The experimental entrainment coefficient ranges between $0.091$ and $0.121$.
These values are consistent in terms of order of magnitude with values reported in the jets/plumes literature for other configurations \cite[e.g.][]{van_reeuwijk_craske_2015,carazzo_kaminski_tait_2006,Koh}. Note that it does not make sense to be more quantitative here, since each configuration gives a different value for $\alpha$.

Furthermore, in the encapsulated graphs of Figure \ref{coldtonguefig}, we plot the normalized radial velocity with the height normalized by the thickness of the jet. The radial velocity is normalized by the maximum values observed within the jet denoted $v_{\mathrm{max}}$, which are determined using the expression:
\begin{equation}
v_{\mathrm{max}}(r)=\mathrm{max}\left[|\left<u(t,r,\varphi,0\leq z\leq b)\right>_{\varphi}|\right],
\end{equation}
This analysis is made for various radii ranging from the edge to the bulk.
For both flux ratios in Figure \ref{coldtonguefig}, all velocity profiles collapse on a unique profile which indicates a self-similar structure.
In addition, when considering a fixed value of $Ra_{\phi}$ and $\Gamma$, a lower flux ratio results in a larger radial extent of the cold jet (not shown).
The propagation of a jet on a heated plate is a well-studied topic \citep{SCHNEIDER_1985,STEINRUCK_1995,higuera_1997,fernandez2014,fernandez2016}, often examined in the context of a jet with uniform inlet velocity on a uniformly heated (imposed temperature) plate.
The extension of the jet is determined by the competition between inertia and buoyancy forces.
The velocity of the jet decreases due to entrainment of surrounding fluid, reducing its inertia, while lateral heating generates a vertical buoyancy force.
The transition point is typically defined when the inertia is balanced by the buoyancy forces, often using a local Froude number to quantify this transition \citep{daniels_1993,fernandez2016}   
\begin{equation}
\mathcal{F}r(r)=\left (\frac{v_{\mathrm{max}}^2(r)}{\beta g \Delta T_{\mathrm{jet}}(r)~b^3(r)}\right)^{1/2} \ .
\end{equation}
Here, $\Delta T_{\mathrm{jet}}$ represents the temperature difference between the bottom and the outside of the cold plume $\left(\left<T\right>_{\varphi}(r,z=0) -\left<T\right>_{\varphi}(r,z=b) \right)$ at a given $r$.
\begin{figure}
    \centering
    \includegraphics[scale=0.4]{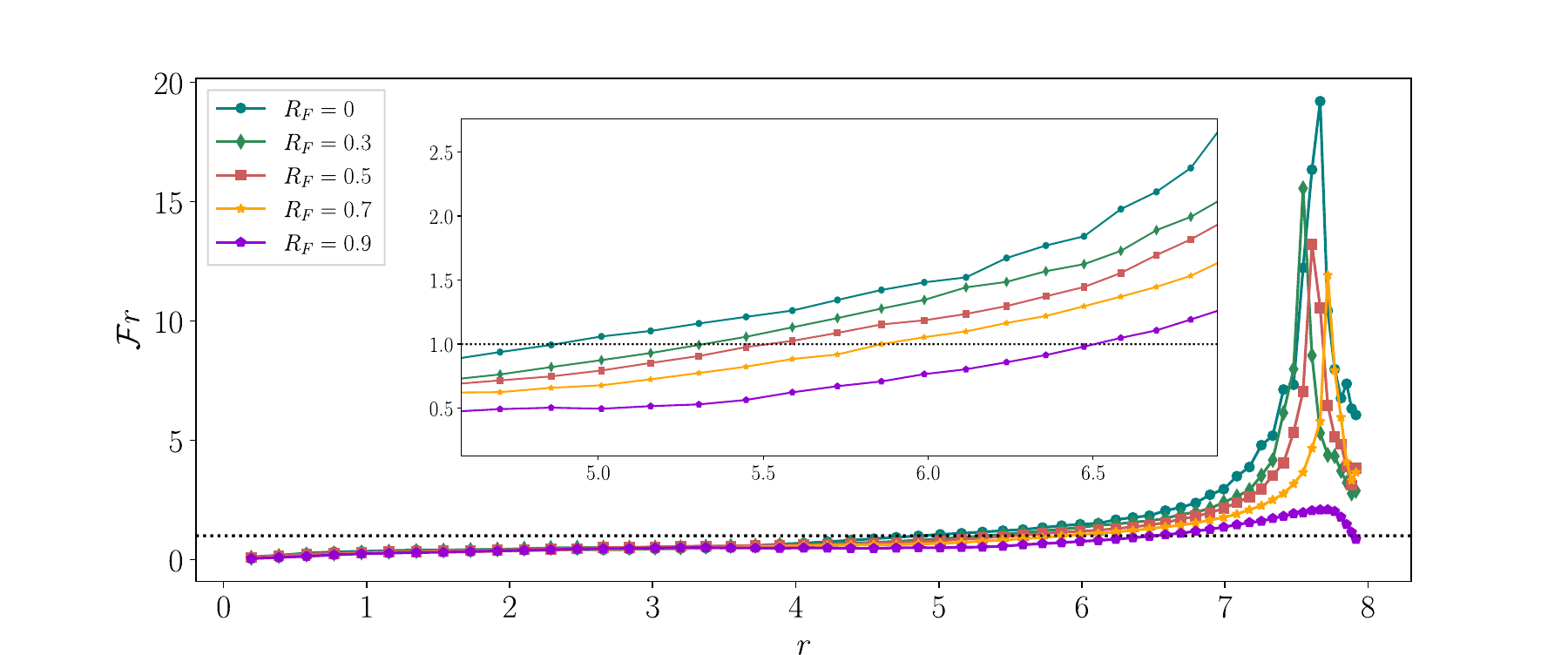}
\caption{Local Froude number as a function of radius for $Ra_{\phi}=10^8$, $\Gamma=8$, $Pr=0.1$ and $R_F$ going from $0$ to $0.9$. The dotted horizontal line indicates $\mathcal{F}r=1$ and the encapsulated plot represents a zoom on the area where $4.5<r<7$.}
    \label{frcoldstrip}
\end{figure}
\
In Figure \ref{frcoldstrip}, the local Froude number is plotted as a function of the radius for several $R_F$ going from $0$ to $0.9$.
Two areas can be distinguished, near the edge where the cold jet inertia dominates ($\mathcal{F}r>1$) and in the bulk where buoyancy finally dominates ($\mathcal{F}r<1$). In order to reach the equilibrium between buoyancy and inertia ($\mathcal{F}r\approx1$), the cold jet travels a greater distance as the flux ratio is lower.
When the Froude number falls below a certain threshold (function of  $Pr$, $\Gamma$, and $Ra_{\phi}$), the cold jet is unable to be maintained, resulting in the radially inward motions being hindered by the adverse pressure gradient caused by buoyancy \citep{daniels_1993,higuera_1997}.
\begin{figure}
    \centering
    \includegraphics[scale=0.3]{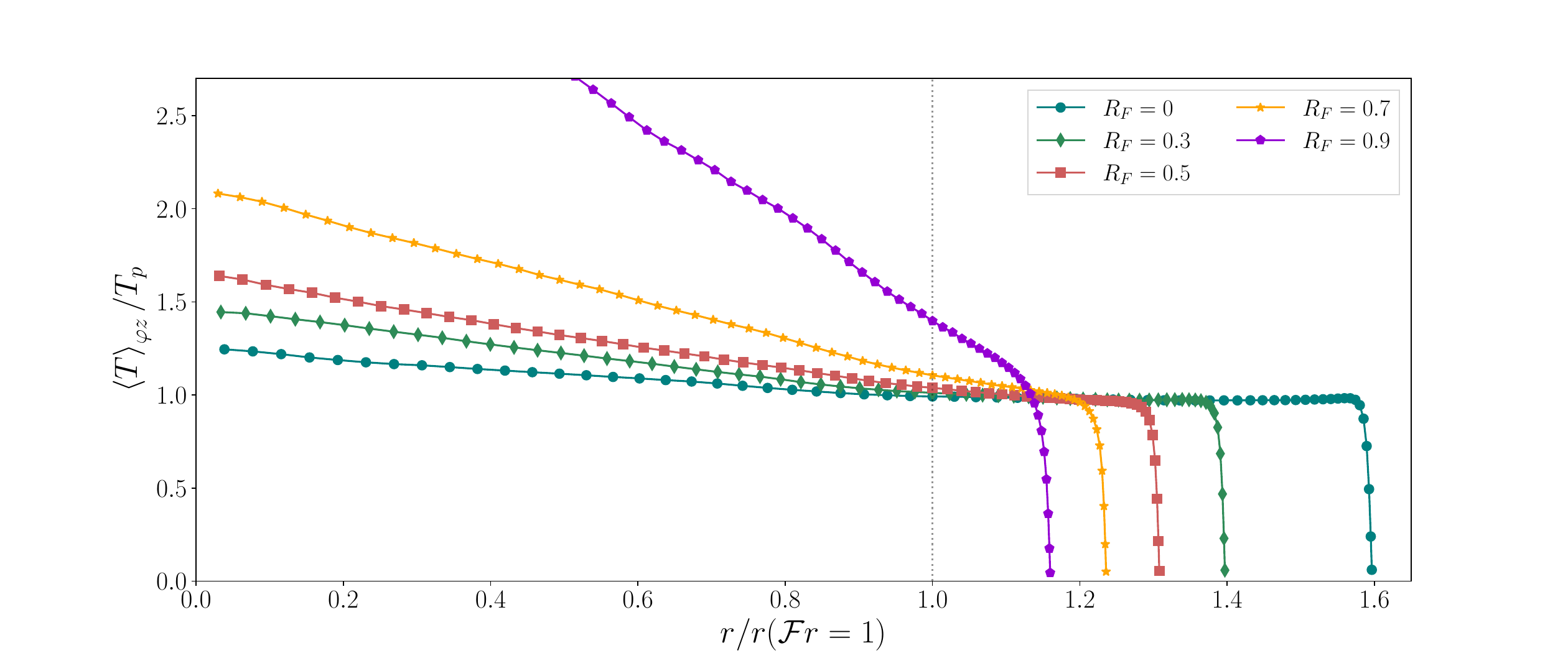}
\caption{Radial temperature profile rescaled by the plateau temperature as a function of the radius rescaled by the radius where $\mathcal{F}r=1$, this for $R_F$ going from $0$ to $0.9$. The input parameters are $Ra_{\phi}=10^8$, $\Gamma=8$ and $Pr=0.1$.}
    \label{Tcoldstrip}
\end{figure}
In Figure \ref{Tcoldstrip}, we plot the radial temperature profile ($\left<T\right>_{\varphi z}(r)$) rescaled by the plateau temperature ($T_p$) as a function of the radius rescaled by the radius where $\mathcal{F}r=1$, for various $R_F$.
We observe that as $r/r(\mathcal{F}r=1)$ approaches 1, the rescaled temperature profiles converge towards the plateau temperature.
This convergence is particularly clear when the flux ratio is low.
As $R_F$ increases, the cold jet region becomes less meaningful, and in the asymptotic case where $R_F=1$, it effectively disappears since all the heat flux is evacuated at the top.
The transition radius in \eqref{eq:1Dmodel} is thus well determined by the parameter $r_p = r(\mathcal{F}r=1)$. We might notice on Figure \ref{Tcoldstrip} that accounting for this threshold is satisfying at first order only: while the critical Froude number is close to 1, its exact threshold value might also slightly depend on the aspect ratio and the Rayleigh number, which is left to future works. 

\section{Conclusions and future works}

In this paper, a systematic numerical study was made of a system where a thin cylindrical layer of fluid ($Pr=0.1$) is heated from below, one part of the heating power being extracted from the top surface, the other part being extracted from the side. This system is inherently heterogeneous in the radial direction: its spatial thermal structure results  from the superposition of two asymptotic regimes corresponding respectively to forced and natural convection. Natural convection is of Rayleigh-Bénard type and is modified by the presence of a convective structure, similar to a turbulent jet, along the bottom surface, originating from the side wall. Combining various scaling laws, we have quantified those two regimes as well as their radial extent to propose a generic model of the radial temperature profile. This 1D model is more relevant than existing models (developed mostly for nuclear safety analyses) which were based only on the average temperature of the system and neglected the radial variations of temperature \cite[see][for more details]{rein}. 

One of the extensions of this work is to relax the constraint of a uniform heat flux at the top and make it dependent on the radial position. This will make the analysis more consistent, since we have shown that the system cannot be considered as homogeneous in temperature, at least in the radial direction. This issue requires careful consideration and systematic analysis, following for instance the recent work of \cite{clarte2021effects}. 

The conclusions of this study allow us to better predict the average quantities of the system. However, relying solely on mean values is insufficient for nuclear safety analysis. Because of the turbulent nature of the system, important fluctuations are observed (in the calculations) at the side wall (see e.g. the observed thermal branches in Figure \ref{low_ra}). Such events cannot be predicted by considering only the mean values. Therefore, it is also necessary to consider the statistics of fluctuations and the potential for extreme values of heat flux.
In this view and for the extreme regimes relevant for the nuclear safety application (i.e. $Ra_\phi$ up to $10^{10}$), the direct simulation tool, even filtered, is limited, in particular to collect enough data for statistical convergence. Hence, our numerical study could be adequately complemented by an experimental study. All the above mentioned points will be the focus of future works.

\section{Declaration of Interests}
The authors declare the following interests regarding the funding and support received for this work: L'institut de radioprotection et de sureté nucléaire (IRSN), Commissariat à l'énergie atomique et aux énergies alternatives (CEA) and  Electricité de France (EDF) provided financial support for this project. The aforementioned organizations had no influence on the design, data collection, analysis, interpretation of results, or the decision to publish. The content of this work remains the sole responsibility of the authors.

\bibliographystyle{jfm}
\bibliography{jfm-instructions}
\newpage
\appendix
\section{Summary of the simulation parameters}\label{appA}
\begin{table}
\centering
\begin{tabular}{ccccccccc}
\noalign{\global\arrayrulewidth=0.01mm}
\multicolumn{1}{c}{$Ra_{\phi}$} & \multicolumn{1}{c}{$\Gamma$} & \multicolumn{1}{l}{$Pr$} & \multicolumn{1}{c}{$R_F$} & \multicolumn{1}{c}{DNS/filtered} & \multicolumn{1}{c}{$\mathcal{E}$} & \multicolumn{1}{c}{$N$} & \multicolumn{1}{c}{$\eta_K /L$}\\ \hline
$10^2$ & $[4,5,8,16]$ & $0.1$ & $0.1$ & DNS & \begin{tabular}[c]{@{}c@{}}{[}2688,3840,\\ 9216,33792{]}\end{tabular} & $6$ & \begin{tabular}[c]{@{}c@{}}{[}14.2,13.4,\\ 8.3,6{]}\end{tabular}\\ 
$10^3$ & $[4,8,16]$ & $0.1$ & $0.9$ & DNS & $[2688,9216,33792]$ & $8$ & $[5.8,4.4,3.1]$\\
$10^3$ & $8$ & $0.1$ & $0.5$ & DNS & $9216$ & $8$ & $3.7$\\
$10^4$ & $[4,8,16]$ & $0.1$ & $[0.1,0.9]$ & DNS & $[2688,9216,33792]$ & $10$ & $[3.3,2.7,2.3]$\\
$10^4$ & $5$ & $0.1$ & $0.1$ & DNS & $3840$ & $10$ & $2.8$\\
$10^4$ & $8$ & $0.1$ & $0.5$ & DNS & $9216$ & $10$ & $2.5$\\
$3.10^4$ & $[4,8,16]$ & $0.1$ & $0.1$ & DNS & $[2688,9216,33792]$ & $10$ & $[2.5,2,1.8]$\\
$10^5$ & $[4,8,16]$ & $0.1$ & $[0.1,0.9]$ & DNS & $[2688,9216,33792]$ & $10$ & $[1.88,1.56,1.52]$\\
$10^5$ & $5$ & $0.1$ & $0.1$ & DNS & $3840$ & $10$ & $1.92$\\
$10^5$ & $8$ & $0.1$ & $0.5$ & DNS & $9216$ & $10$ & $1.52$\\
$3.10^5$ & $[4,8,16]$ & $0.1$ & $0.1$ & DNS & $[2688,9216,33792]$ & $10$ & $[1.43,1.35,1.36]$\\
$10^6$ & $[4,8,16]$ & $0.1$ & $[0.1,0.9]$ & DNS & $[9984,33608,33792]$ & $10$ & $[4,3.3,1.22]$\\
$10^6$ &  \begin{tabular}[c]{@{}c@{}}{[}5,6,\\ 10,12,14{]}\end{tabular}& $0.1$ & $0.1$ & DNS & \begin{tabular}[c]{@{}c@{}}{[}9984,9984,\\ 33608,33792,33792{]}\end{tabular} & $10$ & \begin{tabular}[c]{@{}c@{}}{[}3.86,3.6,\\ 3,2.73,2.46{]}\end{tabular}\\ 
$10^6$ & $8$ & $0.1$ & $0.5$ & DNS & $33608$ & $10$ & $3.22$\\
$3.10^6$ & $[4,8,16]$ & $0.1$ & $0.1$ & DNS & $[9984,33608,33792]$ & $10$ &  $[3,2.5,1.12]$\\
$10^7$ & $[4,8,16]$ & $0.1$ & $[0.1,0.9]$ & DNS & $[9984,33608,36608]$ & $10$ & $[2.2,1.8,1]$\\
$10^7$ & $5$ & $0.1$ & $0.1$ & DNS & $9984$ & $10$ & $2.3$\\
$10^7$ & $8$ & $0.1$ & $0.5$ & DNS & $33608$ & $10$ & $1.74$\\
$3.10^7$ & $4$ & $0.1$ & $0.1$ & DNS & $9984$ & $10$ & $1.8$\\
$3.10^7$ & $[8,16]$ & $0.1$ & $[0.1,0.9]$ & filtered & $[33792,36608]$ & $10$ &  $[1.44,0.78]$ \\
$10^8$ & $4$ & $0.1$ & $[0.1,0.9]$ & DNS & $9984$ & $10$ & $1.27$\\
$10^8$ & $5$ & $0.1$ & $0.1$ & DNS & $9984$ & $10$ & $1.2$\\
$10^8$ & $8$ & $0.1$ & $0.5$ & filtered & $33792$ & $10$ & $1.3$ \\
$10^8$ & $[8,16]$ & $0.1$ & $[0.1,0.9]$ & filtered & $[33792,36608]$ & $10$ & $$[1.33,0.62]$$ \\
$3.10^8$ & $[4,8,16]$ & $0.1$ & $[0.1]$ & filtered & $[9984,33792,36608]$ & $10$ & $[1.1,1,0.45]$ \\
$10^9$ & $[4,8,16]$ & $0.1$ & $[0.1,0.9]$ & filtered & $[9984,33792,36608]$ & $10$ & $[0.94,0.7,0.23]$ \\
\end{tabular}
\caption{Simulations summary (DNS or filtered) according to the physical and numerical parameters.}
\label{table:1}
\end{table}
\end{document}